\documentclass[useAMS,usenatbib]{mn2e}
\usepackage{times}

\usepackage{hyperref}
\usepackage{amssymb}
\usepackage{amsmath}
\usepackage{graphicx}
\usepackage{multirow}
\usepackage{epsfig}
\hypersetup{backref,  pdfpagemode=FullScreen,  colorlinks=true, citecolor=blue}

\newcommand{\HI}{\mbox{H\,{\sc i }}}

\title[Models of Stephan's Quintet]{Models of Stephan's Quintet: Hydrodynamic Constraints on the Group's Evolution}
\author[J.-S. Hwang et al.]{Jeong-Sun Hwang \thanks{E-mail: 
jshwang@kias.re.kr (J-SH); curt@iastate.edu (CS); 
florent.renaud@astro.unistra.fr (FR); apple@ipac.caltech.edu (PA)}$^{1,2}$, 
Curtis Struck$^{1}$, Florent Renaud$^{3,4}$, and Philip Appleton$^{5}$\\
$^{1}$Department of Physics and Astronomy, Iowa State University, Ames, IA 50011, USA\\
$^{2}$School of Physics, Korea Institute for Advanced Study, Seoul 130-722, Republic of Korea\\
$^{3}$Observatoire Astronomique and CNRS UMR 7550, Universit\'e de Strasbourg, 11 rue de l'Universit\'e, F-67000 Strasbourg, France\\
$^{4}$Laboratoire AIM Paris-Saclay, CEA/IRFU/SAp, Universit\'e Paris Diderot, F-91191 Gif-sur-Yvette Cedex, France\\
$^{5}$NASA Herschel Science Center (NHSC), California Institute of Technology, Mail code 100-22, Pasadena, CA 91125, USA}

\begin{document}

\date{\today}

\pagerange{\pageref{firstpage}--\pageref{lastpage}} \pubyear{0000}

\maketitle

\label{firstpage}

\begin{abstract}
We present smoothed particle hydrodynamic models of the interactions
in the compact galaxy group, Stephan's Quintet.
This work is extension of the earlier collisionless N-body simulations of
Renaud et al. in which the large-scale stellar morphology of the group
was modeled with a series of galaxy-galaxy interactions in the simulations.
Including thermohydrodynamic effects in this work,
we further investigate the dynamical interaction history and
evolution of the intergalactic gas of Stephan's Quintet.
The major features of the group, such as the extended tidal features and
the group-wide shock, enabled us to constrain the models reasonably well,
while trying to reproduce multiple features of the system.
We found that reconstructing the two long tails extending from NGC 7319 toward
NGC 7320c one after the other in two separate encounters is very difficult and unlikely,
because the second encounter usually destroys or distorts
the already-generated tidal structure.
Our models suggest the two long tails may be formed simultaneously
from a single encounter between NGC 7319 and 7320c,
resulting in a thinner and denser inner tail than the outer one.
The tails then also run parallel to each other as observed.
The model results support the ideas that the group-wide shock detected in
multi-wavelength observations between NGC 7319 and 7318b and the starburst region
north of NGC 7318b are triggered by the high-speed collision
between NGC 7318b and the intergalactic gas.
Our models show that a gas bridge is formed by the high-speed collision
and clouds in the bridge continue to interact
for some tens of millions of years after the impact.
This produces many small shocks in that region, resulting a much longer cooling time
than that of a single impact shock.
\end{abstract}


\section{Introduction}
\label{sec:1}

In investigating the evolution of an interacting galaxy system,
multi-wavelength observations and dynamical modeling are complementary.
On the one hand, high-resolution observations in different wave bands
reveal important physical quantities and information about various
physical processes occurring in the system.
By interpreting these quantities and the possible causes of the processes,
we can learn about the dynamical state and deduce a plausible interaction
history of the system.
On the other hand, a well-constrained numerical model not only provide
the direct testing of the plausible interactions and help to interpret
observational results, but it also may yield additional constraints
or information such as the halo profiles of the system.
Ever since \citet{Toomre1972} successfully simulated the stellar morphologies
of several interacting galaxy pairs, explaining the formation of their peculiar structures, 
numerical simulations have become
more popular and important means in studying interacting systems.

%fig. 1
\begin{figure*}%[ht]
\centering
\includegraphics[width=15cm]
{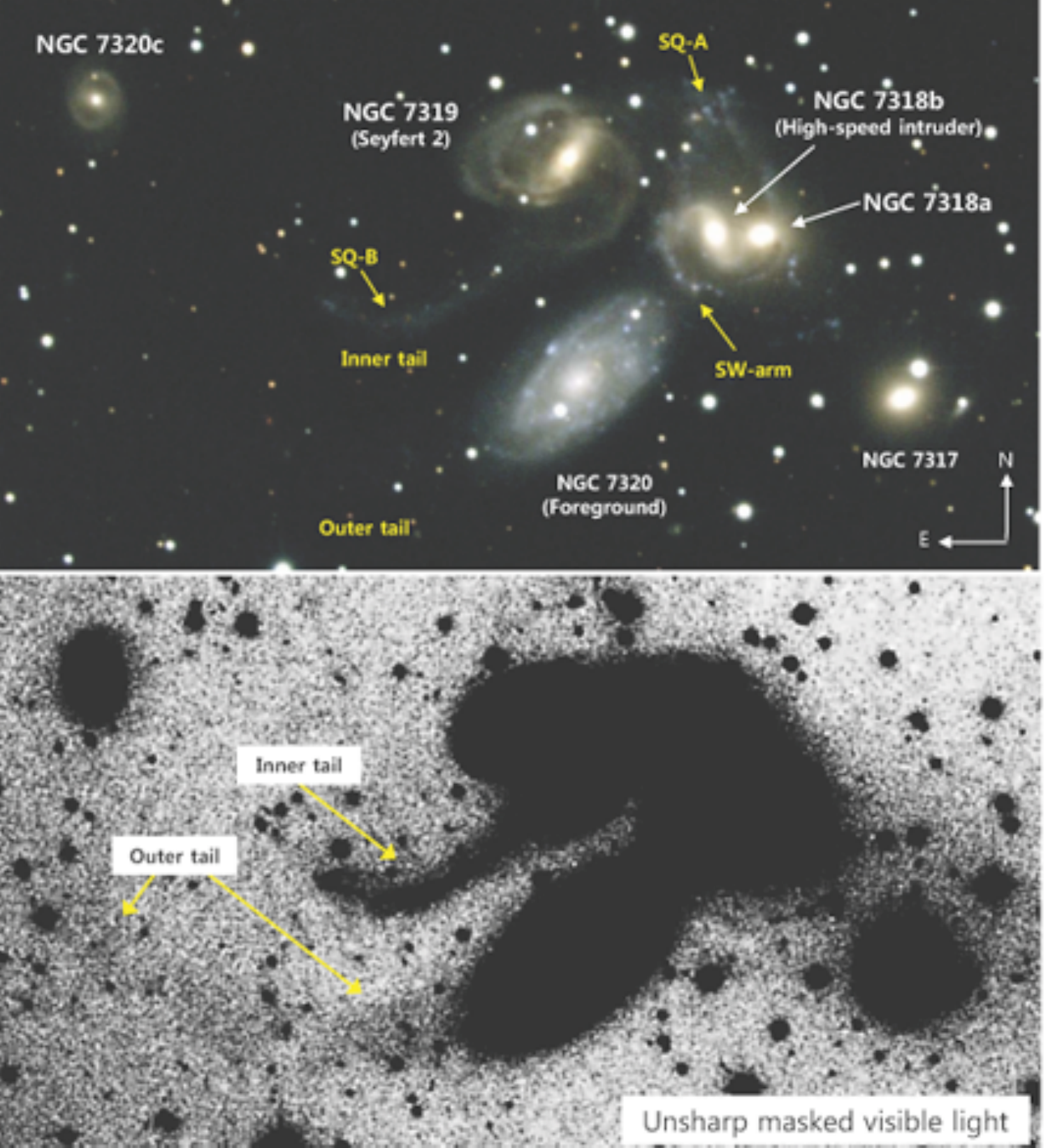}\\
\caption{
\emph{Top}: the optical morphology of Stephan's Quintet.
The members and major large-scale features are marked: the inner
and outer tails (the two long parallel tails; 
the outer tail is hard to see as it is diffuse and passes behind NGC 7320), 
the star-forming regions SQ-A and SQ-B, 
and the tail-like SW-arm feature 
(see text for more descriptions of the features).
NGC 7320 (to the lower middle in the image) is only a
foreground projection.
North is up and east to the left. (Credit: NOAO/AURA/NSF)
\emph{Bottom}: an unsharped-masked version of the above optical image to show 
the outer tail better. The extended outer tail runs parallel with the inner tail.    
\label{fig:1}
}
\end{figure*}

%fig 2
\begin{figure}
\centering
\includegraphics[width=8.0cm]{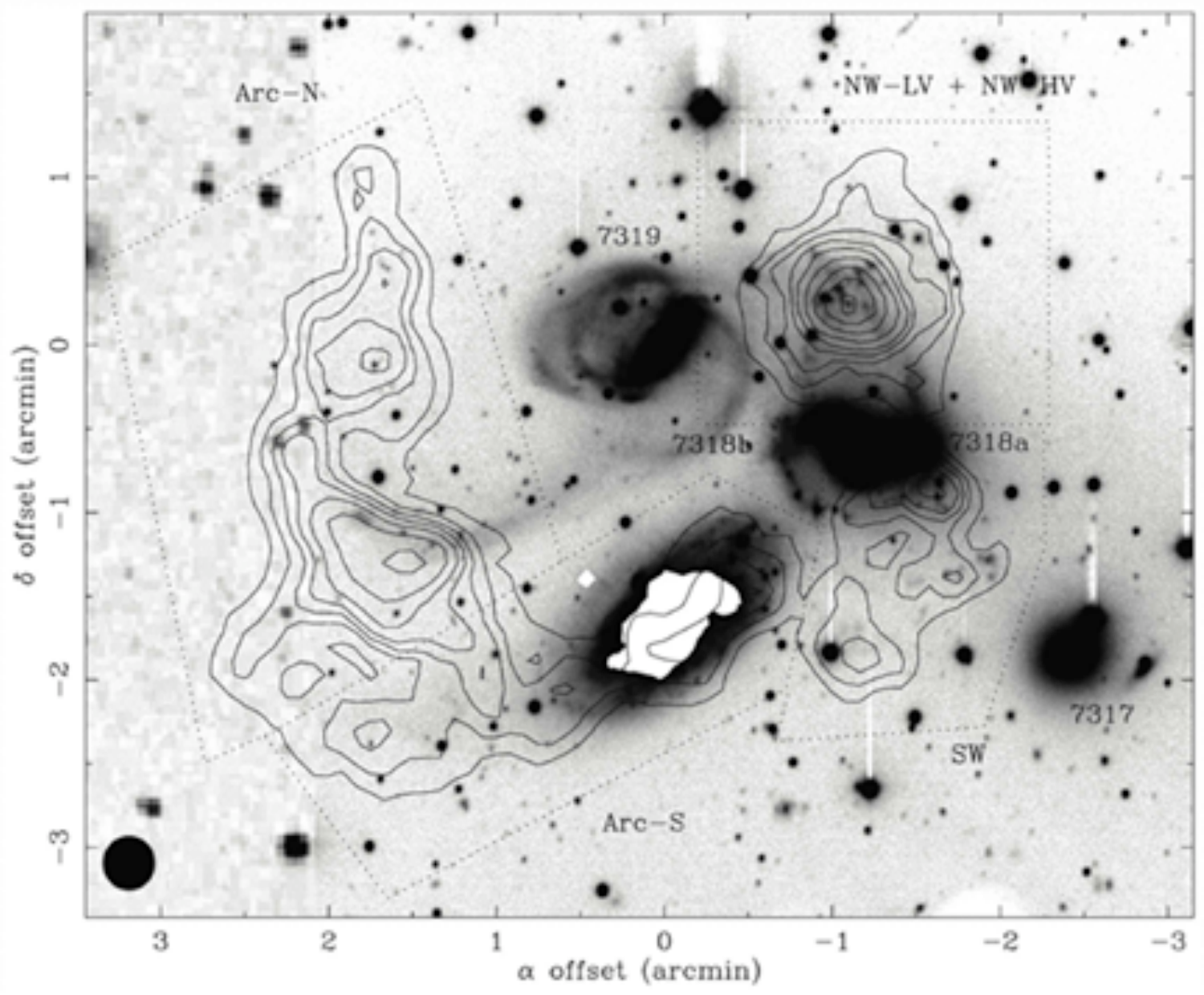}
\caption{
\HI gas observed in Stephan's Quintet (from \citealt{Williams2002}).
Most of the gas is found outside of the galaxies, 
within five physically distinct components and 
in three discrete radial velocity bands: 
Three \HI clouds, `Arc-N', `Arc-S', and `NW-HV',  are detected 
in the highest radial velocity band, 6475$-$6755 km~s$^{-1}$. 
An \HI cloud `NW-LV' is found 
in the intermediate velocity band  5939$-$6068 km~s$^{-1}$,
and a diffuse \HI feature `SW' is detected 
in the low velocity band 5597$-$5789 km~s$^{-1}$. 
\label{fig:2}
}
\end{figure}

%fig 3
\begin{figure}
\centering
\includegraphics[width=7.5cm]{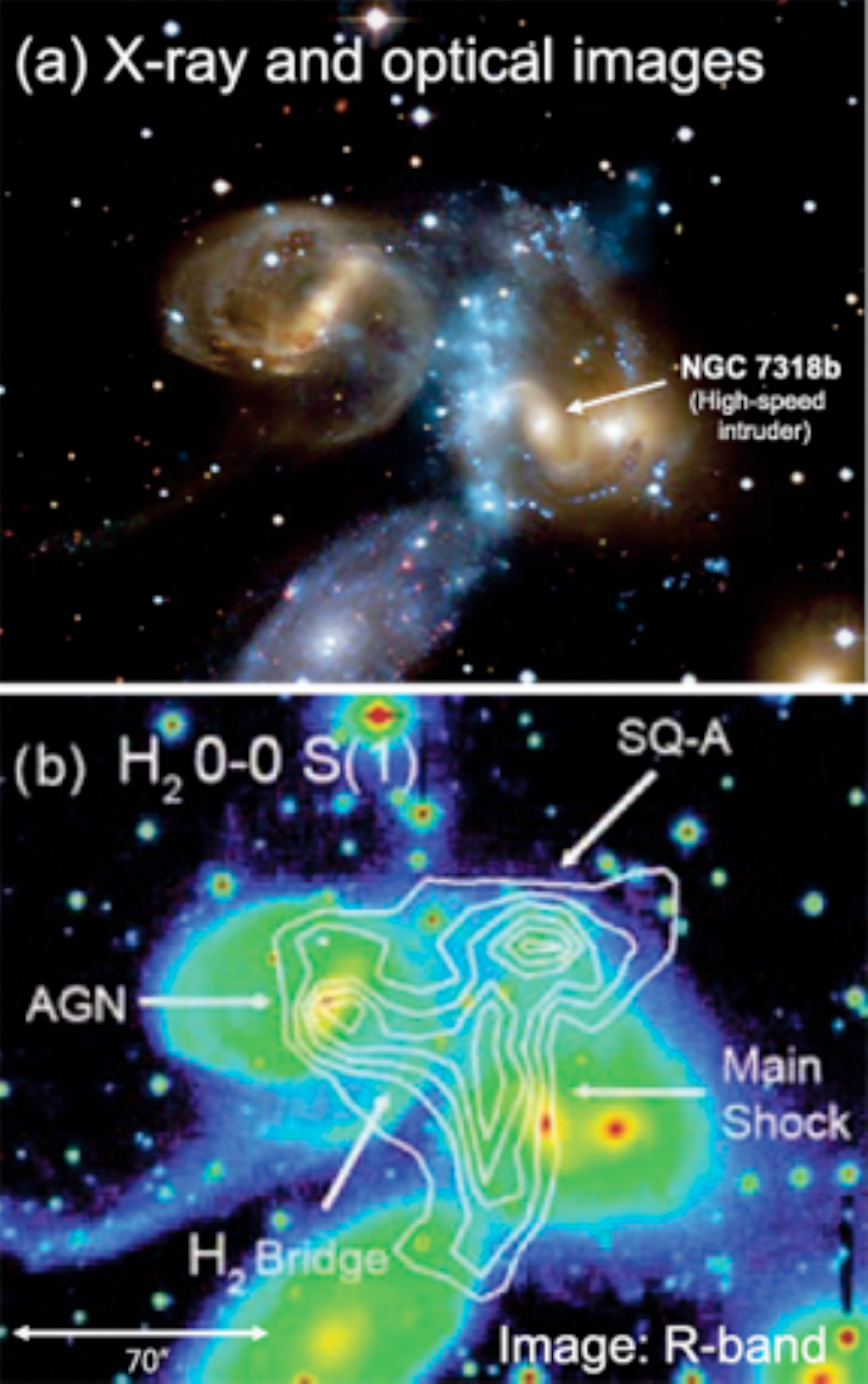}
\caption{
Central region of Stephan's Quintet seen in multi-wavebands.
\emph{(a)} A Chandra X-ray Observatory image (cyan) is superimposed on 
an optical image 
from the Canada-France-Hawaii Telescope. 
A large-scale shock feature appears as a curved, elongated X-ray ridge 
in the middle of the image. 
The shock-heated X-ray gas is thought to be generated by 
NGC 7318b colliding through the core of the group at high speed. 
(Credit: X-ray (NASA/CXC/CfA/E.O'Sullivan); 
optical (Canada-France-Hawaii-Telescope/Coelum))
\emph{(b)} H$_2$ contours (white; from \citealt{Cluver2010})
detected by using \emph{Spitzer Space Telescope}
are overlaid on an $R$-band image (from \citealt{Xu1999}). 
The warm molecular gas shows a similar distribution to that of the
hot X-ray-emitting plasma. 
(North is up and east to the left.) 
The group-wide shock elongated in the north-south direction (`main shock')
and a second feature running eastward (`H$_2$ bridge') are also seen
in H$_2$ emission as in X-ray emission.
\label{fig:3}
}
\end{figure}

Compact groups 
(compact and relatively isolated systems of several galaxies) 
often show highly distorted features of interactions.
Due to the dense environment, galaxy interactions would occur more frequently
in more complex ways in compact groups than binary systems.
These groups are thus important environments for studying various interaction
and merger effects and also the evolution of galaxies 
at higher redshifts where interactions are more frequent (e.g., \citealt{Gottlober2001}).

However, unlike binary systems where computer simulations have often been used
to interpret observations, simulations for compact groups have not been
frequently attempted, especially to model specific groups.
Constraining models is much more difficult or sometimes not doable if the
system has experienced multiple strong interactions involving multiple members and
the structural or kinematical evidence of interactions has been largely erased.

For some special compact groups, however, 
numerical modeling is feasible.
For example, if a system maintains characteristic tidal features or
collisional ring structures then it is possible to limit the plausible
interactions to a few cases. 
The presence of recent interaction is also helpful, 
since there can be unique evidence preserved from the `young' interaction. 
In addition, even though a group appears to have experienced multiple interactions,
if each interaction involved mainly two galaxies rather than three or more members
(i.e., a sequence of `two-at-a-time' interactions),
then modeling interpretations could be more tractable.

Stephan's Quintet (hereafter SQ; also known as HCG 92, Arp 319, and VV 288;
see Fig.~\ref{fig:1}), one of the first compact galaxy groups identified
(\citealt{Hickson1982}), has many of these properties.
It shows strongly disturbed structural and kinematical features, such as
extended tidal tails, disturbed arms and ring-like features, and
rich intergalactic medium (IGM) with little gas remaining in the central region of
every member galaxy (e.g., \citealt{Sulentic2001, Williams2002}; 
see Fig.~\ref{fig:2}). 
Moreover, the group is thought to be experiencing a high-speed collision between
a member (NGC 7318b) and its IGM; multi-wavelength observations,
including radio continuum (e.g., \citealt{Allen1972, Xu2003}) and
X-rays (e.g., \citealt{Trinchieri2003, Trinchieri2005}; see Fig.~\ref{fig:3}(a)),
have consistently shown a huge elongated feature ($\sim$30 kpc)
in the IGM which has been widely interpreted as a shock front triggered by
an ongoing or very recent collision.

More recently, \citet{Appleton2006} discovered strong almost pure-rotational
H$_2$ line emission
along the X-ray-emitting shock front. 
In follow-up observations (\citealt{Cluver2010}; see Fig.~\ref{fig:3}(b)),
the line emission at the main shock region was detected over $\sim$480 kpc$^2$
with a luminosity exceeding that of the X-rays from the shock.
\citet{Guillard2009} modeled the H$_2$ excitation 
in a multi-phased medium 
overrun by a high-speed shock caused by the intruder. 
These models, which dealt with H$_2$ formation and destruction, X-ray emission 
and the survival of dust grains in the shock, showed that 
the H$_2$ power seen in the {\emph {Spitzer}} observations could be explained 
in terms of the dissipation of mechanical energy through turbulence. 
Although these models involve a sophisticated treatment of the micro-physics, 
our current paper attempts to place this work in the context of 
the group dynamics as a whole. 
This coupling between the large-scale gas dynamics and 
the detailed astrophysics of the shocked regions 
provides a strong motivation for this study. 
%This surprising detection of warm molecular gas coexisting with hot gas in the
%shock region motivated us to study the hydrodynamics of the system.
%%In addition, although many observational and theoretical research programs have helped
%%to understand various features in SQ, the evolutionary history and some aspects of the group
%%remain unclear, so there is a need for further detailed simulations to understand the system.

Given all of those interesting features and observational constraints 
available from the literature, 
we have performed smoothed particle hydrodynamics (SPH) simulations of SQ,
taking up to four strongly interacting members into account.
Previously, two of us started modeling of this system with a purely gravitational
N-body code and examined interaction history and large-scale stellar morphology
(\citealt{Renaud2010}).
(Hereafter we distinguish the previous work (code, models, etc) with
the term `N-body'; if not specified, the reference is to the current SPH code, models, etc).
In this paper, we extend the N-body work adding thermohydrodynamic effects
to model the shocks. With new models, we further investigate the origins of the
large-scale features and the dynamical evolution of the intergalactic gas of the system.
In future work, we will examine the star formation history and future evolution
of the group in more depth using SPH models.

In Section~\ref{sec:2}, we first briefly review the members and the large-scale features
characterizing SQ and consider the group's plausible interaction history and model constraints.
Then we explain the simulation code and model details in Section~\ref{sec:3} and
present model results in Section~\ref{sec:4}.
Finally, we summarize and discuss our main results
in Section~\ref{sec:5}.

\section{Overview of SQ properties and model constraints}
\label{sec:2}

\subsection{The members and the large-scale features}
\label{sec:2.1}

The optical morphology of SQ is presented in Fig.~\ref{fig:1}.
By apparent proximity, SQ traditionally denotes the group of five galaxies,
NGC 7317, NGC 7318a, NGC 7318b, NGC 7319, and NGC 7320.
However, NGC 7320 is a foreground galaxy which has a considerably smaller redshift
than others (\citealt{Burbidge1961}) without any indications of interactions with the group.
Actually, NGC 7320c, located far east of NGC 7319, is considered to be linked to the group,
particularly to NGC 7319.
As we conduct numerical modeling of SQ, we treat the five physically related galaxies, 
excluding NGC 7320 and adding NGC 7320c, as the members of the quintet.

SQ is a relatively close
($\sim$94~Mpc away, assuming a Hubble constant of 70 km~s$^{-1}$ Mpc$^{-1}$ and
a systemic velocity of 6600 km~s$^{-1}$ for the group; \citealt{Appleton2006})
and isolated system among the Hickson Compact Groups (HCGs).
Four members, NGC 7317, 7318a, 7319, and 7320c, have a similar recession velocity
of $\sim$6600 km~s$^{-1}$.
In contrast, NGC 7318b has a lower recession velocity than the others by
$\sim$900 km~s$^{-1}$; this galaxy (`high-speed intruder') is thought to be
coming toward us with the large relative velocity, having collided with the material placed
between NGC 7319 and 7318a and triggered 
the group-wide shock shown in Fig.~\ref{fig:3}.

The large spiral galaxy NGC 7319, whose nucleus is known to be a Seyfert 2
(\citealt{Durret1994}), has a strong central bar, disturbed arms, and two long tails
(hereafter the parallel tails, or the inner and outer tails; see Fig.~\ref{fig:1}).
Both inner and outer tails appear in the optical to extend from NGC 7319 toward NGC 7320c,
a small spiral or ringed galaxy.
The outer tail is seen at the south of the inner tail, although the outer one is harder
to see as it is more diffuse and passes behind the foreground galaxy NGC 7320,
running nearly parallel with the inner one.
The inner tail, which is optically brighter than the outer one, is active in star formation.
Many tidal dwarf galaxy candidates have been found along the inner tail
(e.g., \citealt{Hunsberger1996}), including `SQ-B' marked in the figure.
Star formation activity in the outer tail has not been well studied
because the foreground galaxy blocks a large portion of the tail.
In ultraviolet (UV), the inner tail is far more extended toward north than in optical
and shows a loop-like structure 
(\citealt{Xu2005}).

To the west of NGC 7319, a pair of galaxies NGC 7318a/b is seen with complex features
around them, including `SQ-A', and `SW-arm' (Fig.~\ref{fig:1}).
NGC 7318b, the high-speed intruder, has weakly barred spiral morphology and its optical core
looks relatively intact (on the contrary, its interstellar medium (ISM) has been well
removed as shown in Fig.~\ref{fig:2}).
NGC 7318a is the most difficult member for type determination.
Even whether it is an elliptical or a spiral is unclear, due to the complex features around the pair.
Most of the features are thought to be related to NGC 7318b,
however, some of the arm-like structures might be associated with NGC 7318a
(e.g. \citealt{Shostak1984}).
On the other hand, the photometric data of NGC 7318a show a smooth light distribution
like an elliptical (\citealt{Moles1998}).
The galaxy is listed as an E.2.P. in the Third 
Reference Catalogue of Bright Galaxies (RC3; \citealt{deVaucouleurs1991}). 
It was also classified as Sc (\citealt{Hickson1989}) and 
SB0 (\citealt{Hickson1994}).
Features in the north of the pair appear to cross each other, and near the apparent crossing
a starburst region SQ-A has been found (\citealt{Xu1999}),  
together with many young ($<$10 Myr) star cluster candidates (SCCs) 
over a large area (\citealt{Fedotov2011}). 
The feature SW-arm 
looks like a tidal tail, and some H$\alpha$ clumps between NGC 7318a
and 7317 (e.g., \citealt{Xu1999}) 
might be related to SW-arm.

The fifth galaxy NGC 7317 is located to the southwest of the pair NGC 7318a/b.
This elliptical member does not show direct signs of interaction with others
(thus we do not include the galaxy in the models).
However, the diffuse optical and X-ray haloes lying over the regions of NGC 7318a/b and
NGC 7317 (e.g., \citealt{Moles1998})
indicate that the galaxy is a member of the group.

The neutral hydrogen distribution of the system is very interesting.
As shown in Fig.~\ref{fig:2}, an VLA \HI observation
(\citealt{Williams2002})
found most of the gas outside of optical boundaries of the members,
within five distinct components 
(`Arc-N', `Arc-S', `NW-HV', `NW-LV', and `SW')
and in three discrete radial velocity bands 
(6475$-$6755~km~s$^{-1}$, 5939$-$6068 km~s$^{-1}$,
and 5597$-$5789 km~s$^{-1}$; with a Hubble constant of 75 km~s$^{-1}$ Mpc$^{-1}$ 
and a recessional motion of 6480 km~s$^{-1}$ for the center of mass 
of the quartet NGC 7317, 7318a/b, and 7319). 
The gas with the highest radial velocity (6475$-$6755 km~s$^{-1}$) is detected
in two large arc-like clouds (Arc-N and Arc-S) tracing the optical inner and outer tails 
(see fig.~6 of \citealt{Williams2002} for the view of each arc-like cloud separately)
and in a relatively compact cloud (NW-HV) centred near SQ-A.
Another kinematically distinct cloud (NW-LV) centred near SQ-A is observed in the intermediate
velocity band (5939$-$6068 km~s$^{-1}$); this cloud coincides with the high velocity one
but is more extended (see figs~8 and 9 of \citealt{Williams2002} 
to see NW-HV and NW-LV separately). 
The low velocity (5597$-$5789 km~s$^{-1}$) gas is found between NGC 7318a/b and
NGC 7317 in a diffuse feature (SW).

The shock seen in X-rays in the central region of SQ is shown
in Fig.~\ref{fig:3}(a) (cyan),
as a narrow elongated feature in the IGM between NGC 7319 and 7318b.
As noted earlier, it is widely accepted that the hot X-ray-emitting gas
in the elongated feature has resulted from the shock-heating triggered by
the present-day high-speed collision between NGC 7318b and the IGM.
The fact that the main shock occurs in a region where
the cold neutral hydrogen is missing 
also supports this idea.

The white contours in Fig.~\ref{fig:3}(b) represent 
the 0-0 S(1) 17$\mu$m rotational H$_2$ line emission 
detected in the central area of the group
(see also figs~2 and 3 of \citealt{Cluver2010}).
The strong line emission in north-south direction
(`main shock') follows the X-ray emitting shock shown in Fig.~\ref{fig:3}(a); 
strong line emission also comes from the starburst region SQ-A and
from the region associated with the active galaxy NGC 7319.
There is another H$_2$ structure (`H$_2$ bridge')
which runs eastward from the main shock.
The structure is also discernable in the X-ray emission.
Overall, the warm molecular gas shows a similar distribution
to that of the hot X-ray emitting gas.
The line luminosity of the H$_2$ in the main shock was measured
to be about three times stronger than the X-ray luminosity from
the hot shocked gas, implying that H$_2$ lines are a stronger
coolant than X-ray emission (\citealt{Cluver2010}).

H$\alpha$ emission in the central region of the group
has also been detected,
with two different components of the emission,
$\sim$5700 km~s$^{-1}$ (which is in the velocity range of the high-speed intruder)
and $\sim$6600 km~s$^{-1}$(\citealt{Xu1999}; see also \citealt{Xu2003}).
It is shown that the shock feature appears in the 6600 km~s$^{-1}$ component, 
but not in the 5700 km~s$^{-1}$ component; 
this indicates the shock front is associated with the 6600 km~s$^{-1}$ component.

\subsection{Interaction history and model constraints}
\label{sec:2.2}

The parallel tails associated with NGC 7319 
are one of the most characteristic features resulting from past interactions
in the group.
Based on the shape and probable age of each tail, three different formation scenarios of the
tails have been considered: (A) both tails could be generated simultaneously by an encounter
of NGC 7320c with NGC 7319 (\citealt{Moles1997}), (B) the outer and inner tails might be
produced one after the other in at least two passages of NGC 7320c around NGC 7319
(\citealt{Moles1997, Sulentic2001}), and (C) the outer tail might be produced by an encounter
of NGC 7320c with NGC 7319 and then the inner tail by a different encounter of NGC 7318a
with NGC 7319 (\citealt{Xu2005}).

Scenario A implies the same formation ages for both tails; a bridge and a counter tail of
NGC 7319 which were supposed to be pulled out after a closest approach of NGC 7320c
could later develop into the inner and outer tails, respectively.
The outer tail, however, has been often interpreted to be `older' (formed earlier) than
the inner tail as it is optically fainter and broader than the inner one, so this scenario has
not been favored over the other two where the outer tail is formed earlier.
Recently, \citet{Fedotov2011} suggested that 
the inner and outer tails would be formed $\lesssim$~200 Myr 
and $\sim$~400 Myr ago, respectively, 
by the age determination of the SCCs.  
However, we note that although the northern inner tail appears to contain
more recent star formation than the southern outer tail,
this does not preclude the possibility that the two were formed together
but suffered different star formation histories 
(because the outer tail is more farther out (i.e., more spread out or 
in a lower density environment)). 
Moreover, the age determination of the outer tail is estimated by the clusters 
in only one part of the tail (inevitably, a part which is well away from the 
foreground galaxy) and that of the inner tail 
is based on the blue clusters (but it also contains some old clusters 
with age of $\sim$~500~Myr as well; \citealt{Fedotov2011}). 
We also note that the age determination of the outer tail versus 
the inner tail is based on a systematic shift in B$_{438}$ - V$_{606}$ color of 
0.1 magnitudes between the two sets of clusters (\citealt{Fedotov2011}) 
which could potentially results from differences in extinction between 
the two regions (as the authors point out such extinction would have to be a 
uniform screen since the points in both clusters are quite well clustered 
around their respective means--but the proximity of the foreground galaxy to 
the `old' tail makes this at least plausible). 
Thus we consider the formation of both tails together to not be completely ruled out 
by observations, and is still a plausible hypothesis. 
Further observational work will be needed to confirm a strong age difference 
between the two tails.

Scenario B considers common origins, but at different times, for both tails.
It suggests that an earlier passage of NGC 7320c would produce the outer tail
and then the recent passage (considering a bound orbit of NGC 7320c) would produce
the inner tail, because both of the optical parallel tails point to NGC 7320c
but the more diffuse outer tail might be formed earlier.
Until about a decade ago, the radial velocity difference between NGC 7319 and 7320c
had been (incorrectly) thought to be $\sim$700 km~s$^{-1}$. So, assuming a fast encounter,
\citet{Moles1997} estimated the recent encounter might occur
as recently as $1.5 \times 10^8 \,h^{-1}$ yr ago and an earlier encounter at least
5 $\times$ 10$^8 \, h^{-1}$ yr ago or earlier. 
However, later measurements of the radial velocity of NGC 7320c came
out to be nearly equal to that of NGC 7319 (e.g., \citealt{Sulentic2001}), which implies
the recent encounter would be slow rather than fast (assuming the difference between
the transverse velocities of the two galaxies is comparable to that between the radial velocities).
It is thus argued that if NGC 7320c had encountered NGC 7319 producing
the inner tail and moved to its current position then, with a slow passage,
it would take at least $\sim{5}\times{10}^8$~yr (\citealt{Xu2005}), which is much
older than the previously estimated age of 
the inner tail but close to that of the outer tail.

In scenario C the outer tail is also formed earlier than the inner tail, as in the second scenario.
The two scenarios agree on the origin of the outer tail, but not on that of the inner tail.
As a solution to some of the problems raised in the second scenario regarding the age and
non-optical morphology of the inner tail (e.g., the UV inner tail is extended toward north rather
than points NGC 7320c as mentioned in Section~\ref{sec:2.1}),
scenario C suggests an encounter of NGC 7318a might generate
a tidal tail looking like the UV inner tail, and the estimated age of the inner tail by
the encounter would then be about three times younger than that produced by NGC 7320c,
due to the shorter distant between NGC 7318a/7319 than between NGC 7320c/7319
(\citealt{Xu2005}).

As described above, the origins of the inner and outer tails are not yet clearly understood.
However, no matter how the parallel tails have been generated, NGC 7319 would have
undergone one or more strong prograde encounters to have pulled two substantial tails
out of the disc. The system appears to be strongly disturbed during the interaction, as most
of gas in NGC 7319 (and the other involved galaxies) has been stripped off and huge amount
of gas put into the parallel tails (Fig.~\ref{fig:2}).

At the time when the parallel tails were produced, NGC 7318b (the high-speed intruder)
is thought to be far behind the rest of the group and not to be involved in the generation
of the tails. The galaxy seems to enter the group in the relatively recent past and is
currently passing through the material that has been placed west of NGC 7319 with
the large relative radial velocity ($\sim$900 km~s$^{-1}$) triggering the group-wide shock.

There have been some arguments about whether or not NGC 7318b had interacted
with the group (most likely with NGC 7318a)
before the current collision (and after the production of the parallel tails).
The optical core of the high-speed intruder looks still intact,
which leads the idea that the galaxy might be
entering the group for the first time (\citealt{Moles1997, Sulentic2001}).
However, the gas disc of NGC 7318b has already been well stripped and some tail-like
and ring-like features are seen around the pair NGC 7318a/b; these suggest NGC 7318b
had interacted before (perhaps undergone a partial head-on collision with NGC 7318a),
because the removal of outer discs and the development of tidal features occur after
the closest encounter (\citealt{Williams2002}; see also \citealt{Xu2005}).

It is very difficult to understand the origins of each of the features around the pair
NGC 7318a/b clearly, due to the complex interaction involving the high-speed intruder,
perhaps multiple interactions from the (relatively recent) past through the present; the features
might be influenced by some hydrodynamical effects and/or tidal effects.
Particularly, the features at the eastern side of the high-speed intruder would require more
careful interpretations as the arms of the high-speed intruder and the main shock are very close.

The extra-nucleus starburst region SQ-A may be influenced by
some hydrodynamical processes due to the current high-speed
collision, because according to H$\alpha$ data
the star formation in the region is occurring not only in the ISM
of the high-speed intruder but also in the IGM,
6600 km s$^{-1}$ component, and
the 6600 km s$^{-1}$ component would not be detected if
SQ-A is triggered by tidal effects (\citealt{Xu2003}).
It is worth noting that SQ-A lies nearly at the northern end
of the shock ridge seen in radio continuum and at the southern end
there is also a star-forming region (i.e., `7318b-south' in \citealt{Cluver2010})
but in the ridge itself shows little star-formation (e.g., \citealt{Xu2003, Cluver2010}).
Our models can, in principal, address the question of whether
the existence of two major star formation complexes at either end of the shock structure
is a coincidence, or whether they are regions which, unlike the main shock,
have less turbulent conditions more conducive to the onset of star formation.

\section{Numerical models}
\label{sec:3}

We have constructed an SPH code for numerical simulations of SQ.
As mentioned earlier, before this hydrodynamic modeling
two of us used an N-body code, based on primarily NEMO stellar dynamics toolbox
(\citealt{Teuben1995}), and performed about 3000 exploratory runs to test various
interaction hypotheses and examined the gravitational aspects of the group
(see \citealt{Renaud2010} for details).
In this new SPH modeling, all of the relatively successful N-body models have been
re-tested and some additional ideas of interactions have been tried.
We ran about 170 simulations with the SPH code.
In this section, we first explain the SPH code details and then
describe our models including the modeling strategies
and the differences among the models.

\subsection{The SPH code}
\label{sec:3.1}

The SPH code used in this work has been produced by modifying the SPH code of
\citet{Struck1997} which was originally designed for studying collisions between two disc galaxies
(see \citealt{Struck1997} for details),
so that the code can be applied to systems with more than two galaxies.
In the SPH code for SQ, four strongly interacting members NGC 7318a/b, NGC 7319,
and NGC 7320c have been considered, excluding NGC 7317 (and the unrelated foreground
galaxy NGC 7320), since we simulate the group from the generation
of the parallel tails to the near future, and during the time
the effect of NGC 7317 on the group is considered to be small 
(e.g., \citealt{Moles1997}; \citealt{Fedotov2011}).
It is also because we do not have observational constraints
to include the galaxy into the simulations.

Each model galaxy consists of discs containing gas and
collisionless star particles and a rigid dark matter halo.
The elliptical looking member NGC 7318a is also
initialized with discs, but we adjust some initial parameters
(such as decreasing the angular and increasing the random velocities of its test particles)
to model it as an elliptical galaxy.

In the SPH code, hydrodynamical forces are calculated with
a spline kernel on a grid.
A constant smoothing length and fixed unit cells in a grid are used.
Local self-gravity is calculated between gas particles
in adjacent cells.
A standard artificial viscosity formulation and
a simple leapfrog integration method are used.
The large-scale dynamics of interactions are simulated with
a restricted three-body approximation.
The specific form of the halo acceleration of a particle is
\begin{equation}
a = {\frac{G{M_h}}{\epsilon^2}} {\frac{r/\epsilon}{(1+r^2 /\epsilon^2)^{n_h}}
\label{eq:1}},
\end{equation}
where $M_h$ is a halo mass-scale, $\epsilon$ is a core radius,
and the index $n_h$ specifies the compactness of the halo.

Simple treatments of heating and cooling are included for gas particles using
an adiabatic equation of state.
(For pressure calculation, adiabatic equation of state was assumed.
Heating and cooling were calculated separately.
Constant cooling time-scales were used in three temperature ranges
as described in \citealt{Struck1997}.)
Star formation is assumed to occur when the local gas density exceeds a constant
threshold density and the gas temperature goes below another constant threshold temperature.
The density and temperature thresholds are arbitrary and can be normalized to give realistic
star formation rate for the system.
Dynamical friction
is not included for the current models.
We will add the effects of dynamical friction in the future runs;
with which we will study SQ's future evolution.

The code uses the non-inertial reference frame of a `primary' galaxy.
(The model galaxy for NGC 7319 has been chosen to be the primary.) 
All masses and lengths in the code are scaled to the halo mass-scale
($M_h$ in Equation \ref{eq:1}) and the core radius
($\epsilon$ in Equation \ref{eq:1}) in the halo potential of the primary galaxy.
There is a characteristic sound speed in the code; which is used as a unit velocity.
These code units are dimensionless and have been converted to the physical units
after simulations.

There are fixed spatial boundaries imposed in the code.
The boundaries are large enough to include almost all particles
but not too big for efficient computing performance.
A few distant particles reach the boundaries and are excluded
from the simulations in most runs.

\subsection{Model differences and initial conditions}
\label{sec:3.2}

Most of our simulations run from just before the production
of the parallel tails to shortly after the current high-speed collision.
We divided the modeling efforts into three major stages,
each with a specific goal: (1) to reproduce the inner
and outer tails accurately, stripping much of gas
off the involved members, (2) to test the occurrence of any
intermediate interaction before the present, especially between
the high-speed intruder and NGC 7318a, and (3) to make a
high-speed present-day collision between NGC 7318b and
the IGM of the group.
(These three stages do not have strict boundaries and can overlap;
they are more like conceptual categories, though several interactions
in the group do indeed seem to occur sequentially.)
We will henceforth refer to the model galaxies for NGC 7319, 7318a, 7318b
(the high-speed intruder), and 7320c as G1, G2, G3, and G4, respectively
(from the most massive model galaxy to the least massive one)
to avoid confusion between the model galaxies and the real ones.

Various interactions have been attempted at each stage.
In particular, we spent more time 
at the first stage, trying many different models, than the other two, 
in order to generate the two tidal tails (which are supposed to develop as the parallel tails) 
in the correct configuration out of the disc of G1. 
The parallel tails are one of the most distinctive features resulting from the past
interactions in the group, so a good model of SQ should be able to reproduce them,
and the outcome of the first stage would directly affect the interactions at the later stages.
The models attempting to reproduce the parallel tails at the first stage were evaluated
depending on how reasonably they reproduced the two tails, and those models that
successfully generated the tails were evolved further.
Once we had any complete model that reproduced the overall morphology of the system
relatively well throughout the entire stages, we then made several similar models changing
the values of a few model parameters and ran them again to determine how the models
are affected by particular parameter values.

We tried many models to produce the parallel tails 
as suggested in both scenarios A (with one encounter of G4) and C (with an encounter of
G4 and then an encounter of G2), but not in scenario B (with at least two encounters of G4)
as the previous N-body simulations (which included the effects of
dynamical friction; \citealt{Renaud2010}) already found that scenario B does not work properly.
(We will hereafter describe complete models as `models A' or `models C' 
depending on which model scenario is referred to 
for the production of the inner and outer tails.)
The reason why the N-body models with scenario B failed was: after one tail
(the outer tail) had been produced by the earlier encounter, G4 had to encounter G1
again \emph{very closely} to be able to generate another tail (the inner tail),
but then G4 could not leave G1 far enough toward its current position (actually, G4
collapsed into the larger member G1 in most cases) due to the strong gravitational
attraction and the dynamical friction.

In the N-body simulations, one of models C was chosen as the best model in
reproducing the general stellar morphology of SQ. In the SPH simulations, however,
models A came out generally better than models C.
The differences between the N-body and the SPH modeling are primarily due to the
different halo potentials used in each simulation code and besides, our models are
evaluated not only on the basis of the stellar, but also the gas features.
We will explain more about models A and C, including the advantages and disadvantages
of each type, in the next section when we present the model results.

We finish this subsection by stating initial parameters used in
one of our best model~A runs (the fiducial model).
Many details of the fiducial model are summarized
in Table~\ref{table:1}.
The values in the table were derived from observational data
or modeling experiences with many trials.
Those values are presented in physical units. 
To scale the SPH code, the length and velocity code units have been converted to
1.0~kpc and 5.0~km~s$^{-1}$ in all models,
then the time code unit becomes 200~Myr.
The mass code unit, which is the halo mass-scale $M_h$ for G1 (see Equation \ref{eq:1}),
has been scaled to $4.7 \times 10^{10}$ M$_{\odot}$
in most models including the fiducial model; $M_h$ for G2, G3, and G4 relative to
that of G1 have been chosen to be 0.78, 0.62, and 0.22, respectively.
$\epsilon$ and $n_h$ in Equation \ref{eq:1} for
all members have been set to 1.0 kpc and 1.4 in the fiducial model.
There has been a halo cut-off radius assigned for each member in the fiducial model as
135 kpc (G1), 55 kpc (G2), 80 kpc (G3), and 45 kpc (G4);
which is five times larger than either the gas disc radius (for G1, G3, and G4) or
star disc radius (for G2, which has been set with a smaller gas disc than its star disc).
The halo mass, with the given potential (Equation \ref{eq:1}), for each member out to
the cut-off radius, in the unit of $10^{10}$ M$_{\odot}$, is
12.6 (G1), 8.2 (G2), 7.1 (G3), and 2.4 (G4). No group halo has been assigned.
The initial gas disc of each member in the model, from G1 through G4 in order,
were set to 27~kpc, 8~kpc, 16~kpc, and 9~kpc,
and the initial stellar disc to 18~kpc, 11~kpc, 11~kpc, and 7~kpc.
So, the stellar and gas disc sizes of the least massive member G4
were chosen to be about one third of those of the most massive member G1.
G2 and G3 were selected to be comparable in mass and stellar disc size,
but not in gas disc size. G2 was set to have smaller gas disc with
a lot fewer gas particles compared to G3.
In all models, all galaxies were initialized in the $x$-$y$ plane
(which was chosen to be equivalent to the plane of the sky)
with disc spins in either
clockwise or counter-clockwise directions,
and then rotated around any of the three orthogonal axes as necessary.
In the fiducial model,
G1 and G4 were set with counter-clockwise directional spins,
while G2 and G3 with clockwise directional spins.
All members in the fiducial model were not tilted in any direction.
(The members may not be all exactly face-on, but tilts were ignored
in the fiducial model; different tilts particularly for G3,
the high-speed intruder,
were tested in several similar models to the fiducial model
for comparison).

%table 1
\begin{table*}
\centering
\begin{minipage}{145mm}
\caption{Initial parameters of the fiducial model}
\begin{tabular}{@{}lccccl@{}}
\hline \hline
& G1  & G2 & G3& G4\\
\hline
Halo masses\footnote{The given halo mass is that contained
within a halo cut-off radius.
Gas and star disc masses are negligible in this model; see text.}
({$ \times 10^{10}$ M$_{\odot}$})\footnote{Physical units are used
in this table. Conversion from code units is described in text.}
&12.6 & 8.2 & 7.1 & 2.4 \\
Halo cutoff radii\footnote{In this model, the halo cut-off radii for G1, G3,
and G4 are set to five times their gas disc radii; for G2, five times its star disc radius.
No group halo is applied.}(kpc)
& 135.0 & 55.0 & 80.0 & 45.0\\
Gas disc radii (kpc)
& 27.0 & 8.0 & 16.0 & 9.0\\
Stellar disc radii (kpc)
& 18.0 & 11.0 & 11.0 & 7.0 \\
Gas particle numbers
& 68,680 & 6,000 & 24,000 & 7,480 \\
Star particle numbers
& 32,000 &  11,760 & 11,760 & 4,960 \\
Disc orientations\footnote{All galaxy discs are initialized in the
$x$-$y$ plane with counter-clockwise directional spins
as seen from the positive $z$-axis and then rotated as necessary.
In this model, G1 and G4 are set in the $x$-$y$ plane with counter-clockwise
directional spins, and G2 and G3 are flipped around the $x$-axis so that they have
clockwise directional spins.
No additional tilts are applied in this model.}
& &$180^{\circ}$ about $x$-axis
& $180^{\circ}$ about $x$-axis & \\ \\
Initial centre positions\footnote{Coordinates are defined in a
conventional, right-handed frame, with the origin fixed at the centre
of the primary, G1.
The $x$-$y$ plane is defined as the plane of the sky,
and positive $z$ as the direction towards the observer.}
& at origin & ($-$70.0, 10.0, $-$20.0)
& (12.0, 2.0, $-$340.0) & (12.5, $-$15.3, 15.3)\\
($x$,$y$,$z$) (kpc)\\
Initial centre velocities
\footnote{Positive $z$ velocities mean  motions towards us.}
& (0.0, 0.0, 0.0) & (110.0, $-$27.0, $-$72.5)
& (20.0, $-$7.5, 300.0) & (35.9, 79.5, $-$77.5) \\
($v_x$,$v_y$,$v_z$) (km s$^{-1}$)\\
\hline
\label{table:1}
\end{tabular}
\end{minipage}
\end{table*}

\section{Simulation Results}
\label{sec:4}

In this section we present the results of our models.
We first describe the general evolution of the fiducial model,
as a representative for models A, focusing on the generations of
the large-scale structural features in the system.
Then we compare our models A to models C, discussing the
advantages and disadvantages of the models.
Finally, we analyse the gas properties in the fiducial model
and compare to observations.

\subsection{Evolution of the fiducial model}
\label{sec:4.1}

\begin{figure*}
\centering
\includegraphics[width=5.8cm]{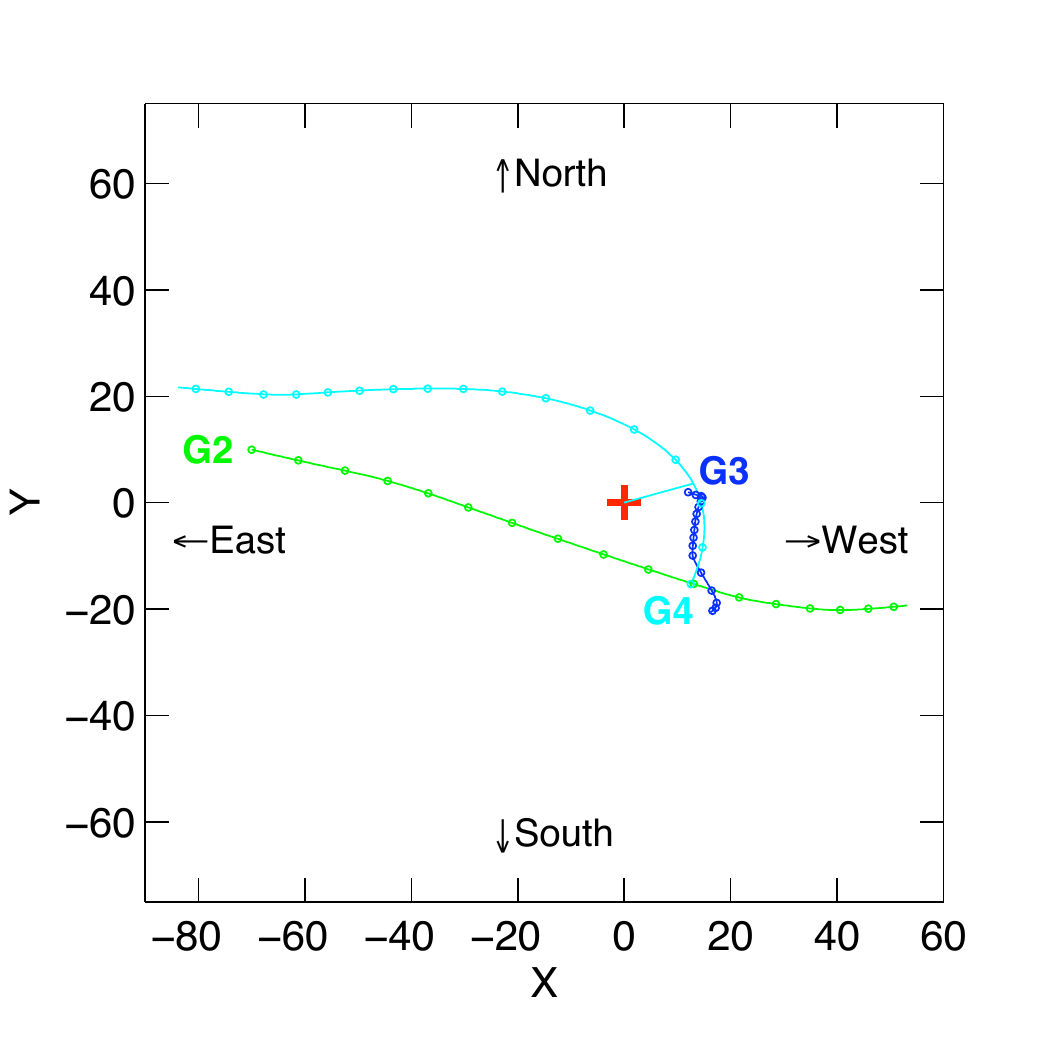}
\includegraphics[width=5.8cm]{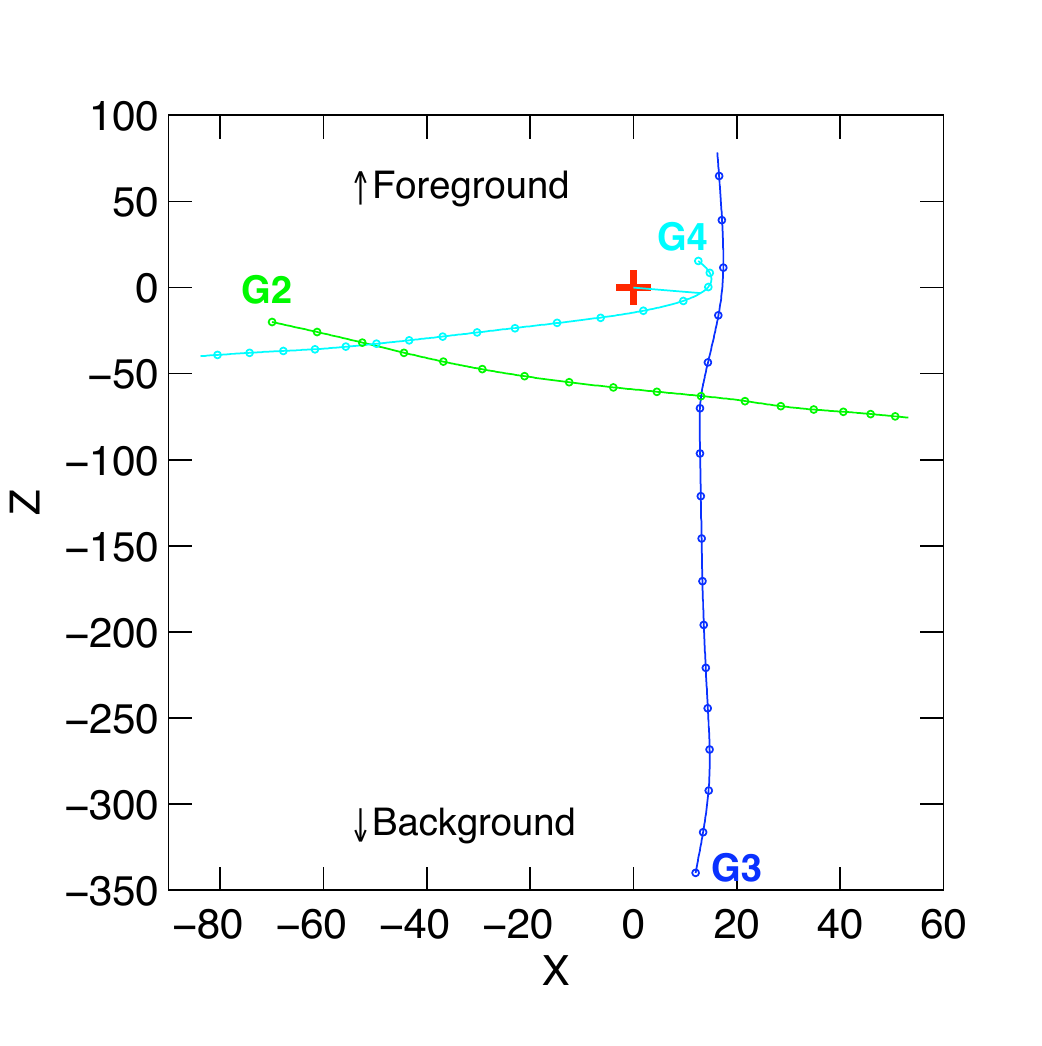} %\quad
\includegraphics[width=5.8cm]{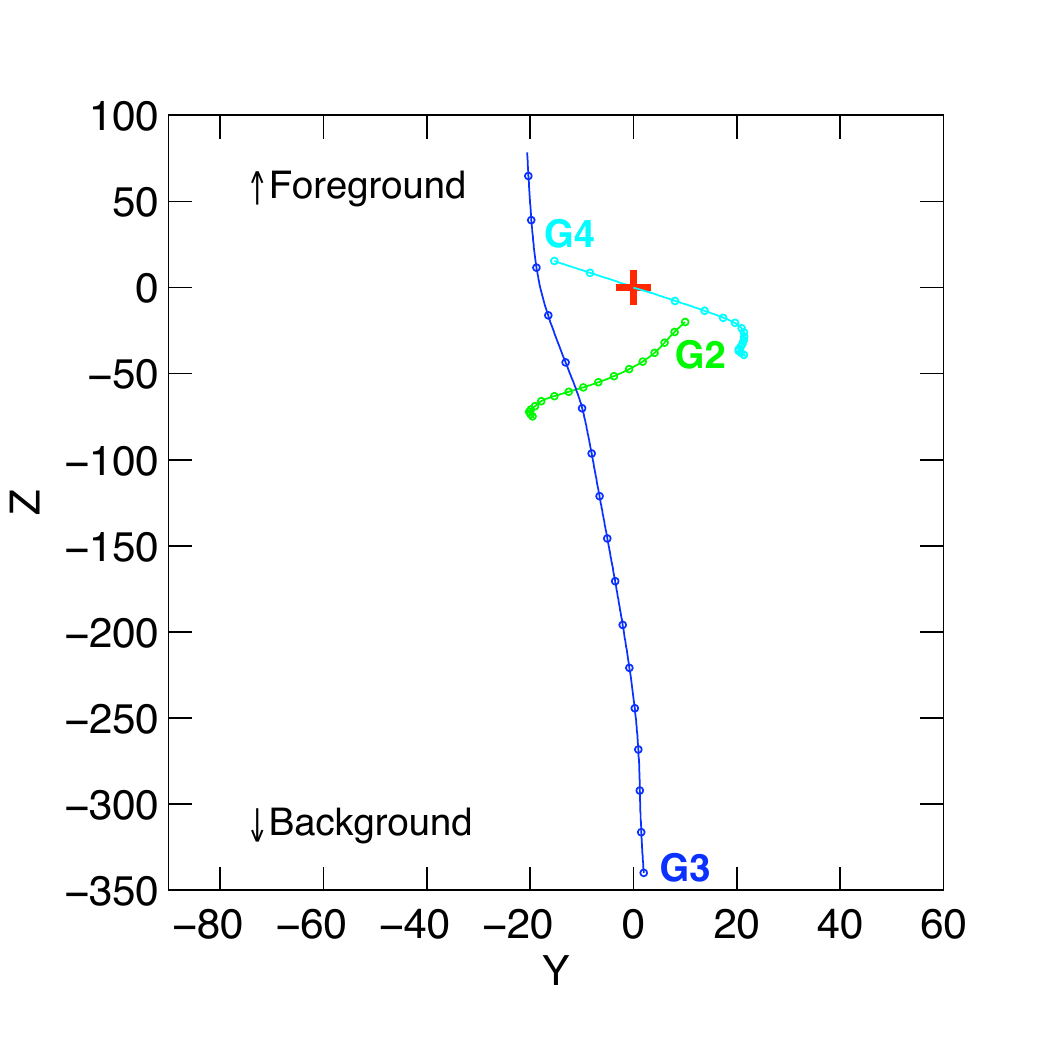}
\caption
{
The trajectories of the four model galaxies in the fiducial model.
The left, middle, and right panels show 
the trajectories projected on to the $x$-$y$ plane (i.e. plane of the sky),
the $x$-$z$ plane, and the $y$-$z$ plane, respectively.
Red, green, blue, and cyan are used to represent the centre positions
of the halo potentials of G1, G2, G3, and G4, respectively.
G1 is fixed at the origin (marked with a red plus sign)
throughout the simulation.
Near the initial positions of the other three members
(see also Table \ref{table:1}),
the model galaxy names (G2, G3, and G4) are indicated.
Little knots marked in the orbital trajectories of G2, G3, and G4 represent
the positions of the model galaxies at every 0.4 time unit
($\sim$ 80 Myr in the representative scaling).
The closest approach between G1 and G4 is indicated with a cyan line.
The axes are marked in code length units
($\sim$ 1 kpc in the representative scaling).
\label{fig:4}
}
\end{figure*}

\begin{figure*}
\centering
\includegraphics[width=14.5cm]
{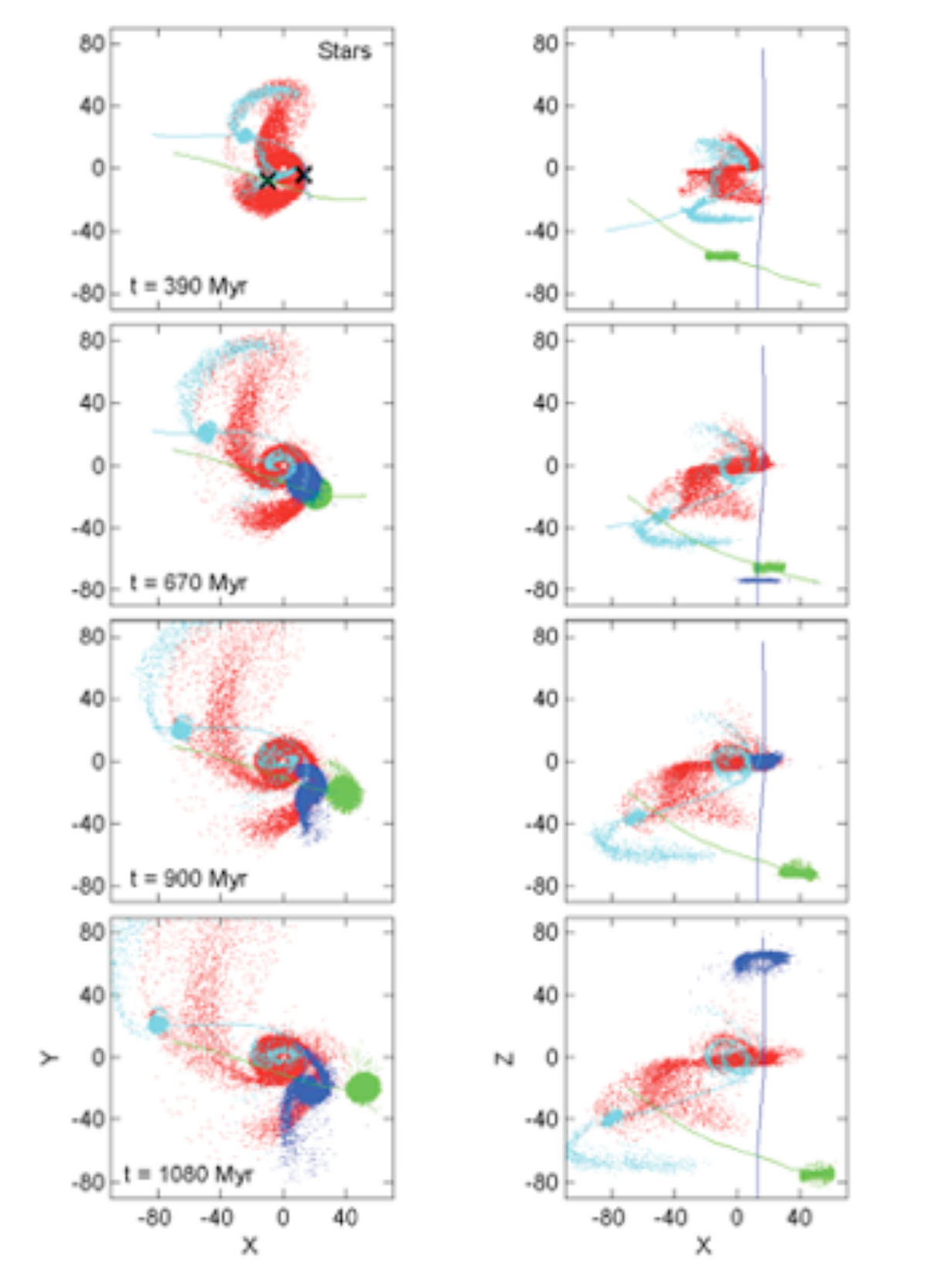}\\  [-20pt]
\caption{
Four snapshots from the fiducial model of star particles projected
on to the $x$-$y$ plane (left column) and the $x$-$z$ plane (right column).
In all panels, particles originating from G1, G2, G3, and G4
are plotted with red, green, blue, and cyan dots, respectively
(the red dots were plotted at first and then green, blue, and cyan dots
in order).
The orbital trajectories (Fig.~\ref{fig:4}) of the members
are overlaid in the same colors used for the particles.
Time indicated in the left panel of each row is measured from the closest approach
between G1 and G4.
In the left panel of the first row, the star particles of G2 and G3 are not displayed
in order to show the bridge and counter-tail pulled out of G1 shortly
after the encounter with G4 more clearly
(instead, the centre positions of G2 and G3 are indicated with `X's).
In the right panel of the top row, G3 is not shown as it is at z $\sim-$ 163.
The second row is at the onset of the collision between G2 and G3.
The third and bottom rows are at times near 
the present, when G3 is passing the mid-plane of the disc of G1 
colliding with the particles placing between G1 and G2, 
and when G3 has passed through the plane of G1, respectively.
\label{fig:5}
}
\end{figure*}

\begin{figure*}
\centering
\includegraphics[width=14.5cm]
{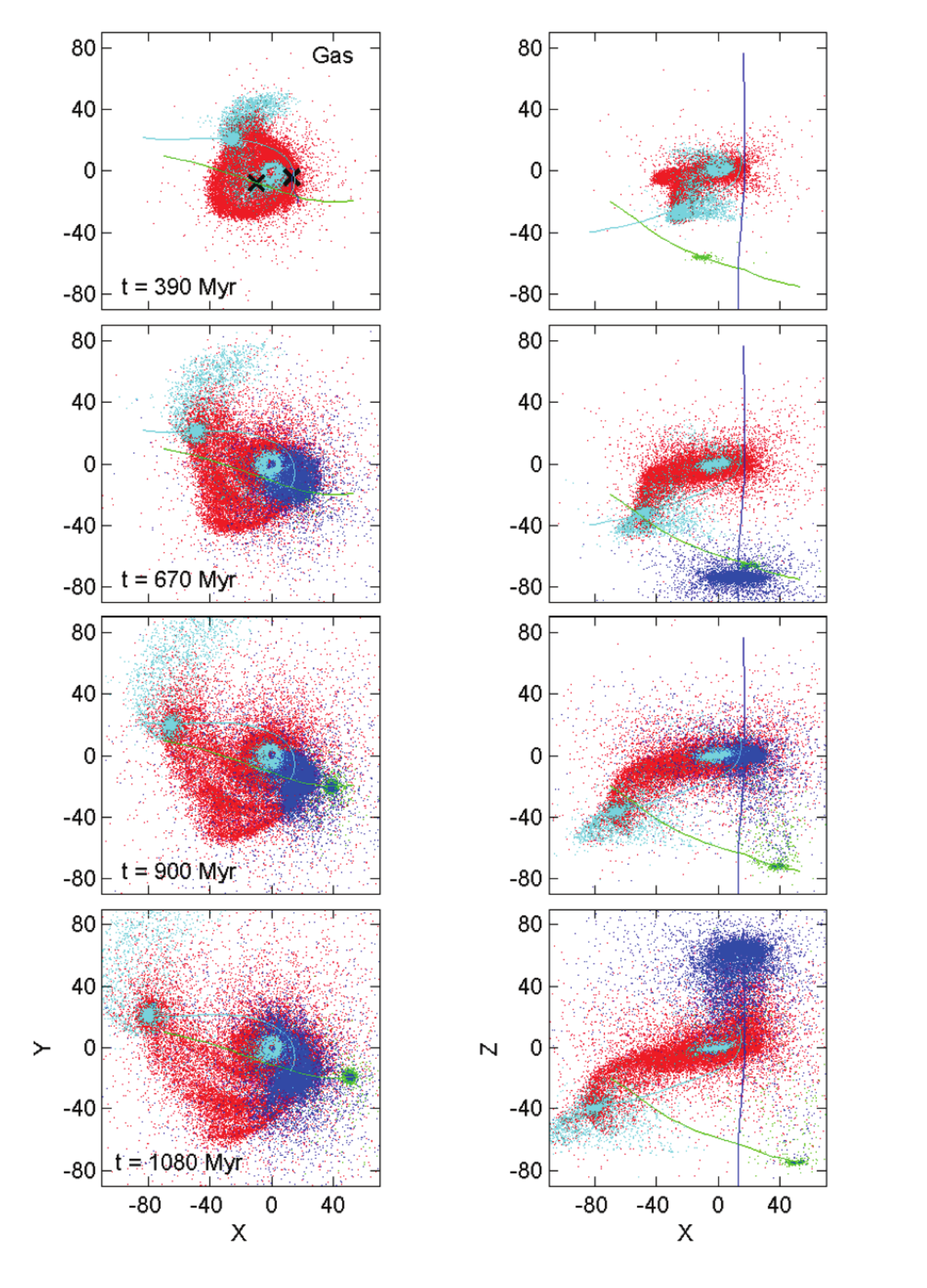}\\  [-20pt]
\caption
{
Four snapshots from the fiducial model of gas particles projected
on to the $x$-$y$ plane (left column) and the $x$-$z$ plane (right column).
These snapshots are taken at the same times as those in Fig.~\ref{fig:5},
showing the distribution of gas particles.
Colors are used in the same ways in Fig.~\ref{fig:5} (red for G1,
green for G2, blue for G3, and cyan for G4).
In the left panel of the first row, the gas particles of G2 and G3 are not displayed
in order to show the early configuration of the bridge and counter-tail pulled out of G1
from the encounter with G4 more clearly
(instead, the centre positions of G2 and G3 are indicated with `X's).
In the right panel of the top row, G3 is not shown as it is at z $\sim-$ 163.
The second row is at the onset of the collision between G2 and G3.
The third and bottom rows are at times near the present.
At the time of the third row, G3 is passing the mid-plane of the disc of G1, 
making a strong impact colliding with 
many gas particles west of G1. 
By the the time of the bottom row, G3 has passed through 
the plane of G1 above $\sim$ 65 kpc;
there are still many gas particles are seen between G1 and G3 
(in the $x$-$z$ view)
forming a gas bridge.
\label{fig:6}
}
\end{figure*}

\begin{figure*}
\centering
\includegraphics[width=13.5cm]
{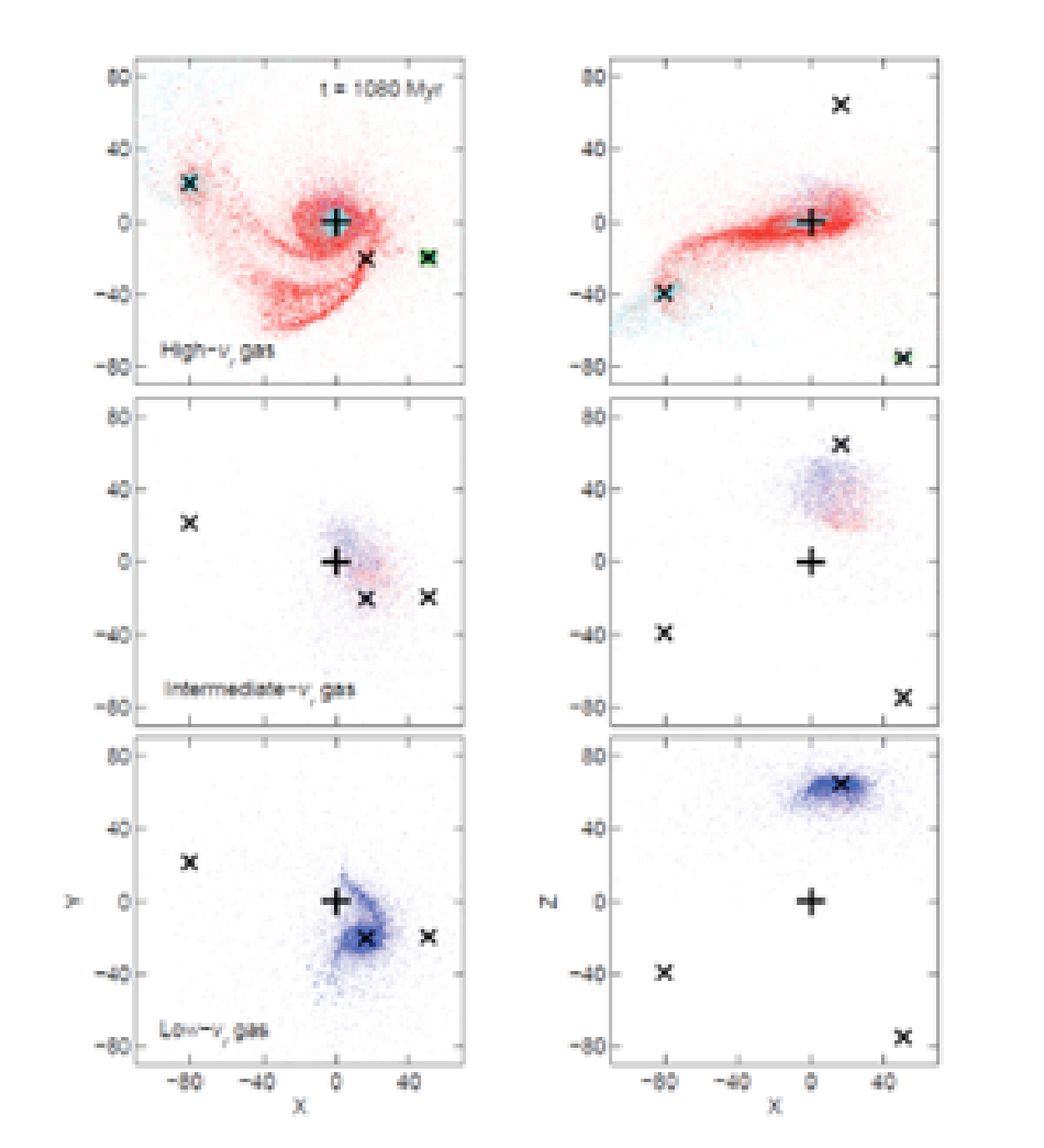}\\ [-15pt]
\caption
{
Gas particles of the fiducial model at $t$ = 1080 Myr 
in three $v_r$ ranges.
The top to bottom rows show the gas particles in high,
intermediate, and low $v_r$ ranges (see the text how to divide the three ranges).
The left and right column show the gas particles projected on to the
$x$-$y$ plane and $x$-$z$ plane, respectively.
Red, green, blue, and cyan dots represent the particles originating from
G1, G2, G3, and G4, respectively (red dots are plotted first and then green,
blue, and cyan in order; the dots are  expressed smaller in this figure than 
in Fig.~ \ref{fig:6} ).
The centre position of G1 is indicated with a plus sign and 
those of G2, G3, and G4 with `X's.
\label{fig:7}
}
\end{figure*}

The fiducial model is designed firstly to generate the inner and outer tails simultaneously
by a close encounter of G4 with G1 (as suggested in scenario A in Section \ref{sec:2.2}),
then at the second stage to have a collision between G2 and G3 relatively far behind 
the plane of G1, and finally at the third stage to make a high-speed collision between G3
and the material found west of G1
(see Section \ref{sec:3.2} for the description of the modeling stages).
The initial parameters and the subsequent orbits of the model galaxies are presented in
Table~\ref{table:1} (and also in Section \ref{sec:3.2}) and Fig.~\ref{fig:4}, respectively.
The distributions of star and gas particles of the model galaxies at four times are shown
in Figs~\ref{fig:5} and \ref{fig:6}, respectively.
(See also Fig.~\ref{fig:7} for a better view of the gas distribution from the model 
near the present, displaying it in three discrete velocity bands separately.)  

%fig 4

%fig 5

%fig 6

As mentioned earlier, we use the coordinate system that
moves with the halo centre of G1
(the most massive and largest member in the model)
so the galaxy appears fixed at the origin throughout the simulations
and set the $x$-$y$ plane as the plane of the sky.
The discs in the fiducial model are initialized with star and gas particles as described
in Table \ref{table:1}.
The adopted scaling is 1~kpc for the length unit, 5~km~s$^{-1}$ for the velocity unit,
and 200~Myr for the time unit in all models.
Note, however, that this (representative) scaling is not unique; 
can be scaled,
e.g., with the total system mass.

The fiducial simulation starts shortly before G4 makes its closest approach to G1.
As shown in Fig.~\ref{fig:4}, G4 starts southwest vicinity of G1 seen in the sky plane,
swings closely around G1 in a counter-clockwise direction, and moves toward east.
Seen from the sides, G4 starts a little bit above G1, turns quickly around G1 (penetrating G1)
going downward, and then slows in the vertical ($z$) direction as it moves toward east.
On the other hand, G2 starts far from G1 and moves toward west going deeper behind G1.
(This orbit of G2 has been chosen in such a way that G2 has little to no effect to the
development of the parallel tails, which were supposed to be generated by an encounter of G4
with G1, staying relatively far from both tails until they grow well, and then G2 can encounter
with G3 well below the plane of G1). G3, the high-speed intruder, has been started far below
all the others, set to move nearly vertically (in $z$-axis) toward the rest of the group,
meet with G2 on its way up, and then pass through between G1 and G2 at a high speed.

The top rows of
Figs~\ref{fig:5} and \ref{fig:6}
show the early development of the inner and outer tails,
in the $x$-$y$ view (left panels) and the $x$-$z$ view (right panels),
at about
$t$ = 1.95 units or 390~Myr (in the representative scaling)
after the closest encounter of G4 with G1.
(We measure time from the instant of the closest approach
between G1 and G4 throughout this paper for convenience;
the start of the model is at about $-$1.0 in code time units. 
As we noted earlier, 
the scaling is not unique; there could be up to 
a factor of two or so change possible in the scaling.) 
By this time, G4 has swung around G1 in a counter-clockwise direction
passing through the large disc of G1 from front to behind
and has pulled two massive arms (a bridge and a counter-arm)
out of the disc of G1.
Since the initial disc spin of G1 was also set
in counter-clockwise direction (Table \ref{table:1}),
G1 feels the encounter as prograde.
(For the initial disc spin of G4, we have tried in both 
counter-clockwise and clockwise directions.
With the counter-clockwise spin, as in the fiducial model,
G4 also feels the encounter as prograde.
In this case, much more gas is scattered to the northeast than the other case;
with sensitive enough observations, this could be used to determine whether
G4 experienced the collision as prograde or retrograde.)
The early encounter between G1 and G4 in this model is somewhat similar to that between
M51 and NGC 5195 proposed by \citet{Toomre1972} in their numerical model.
In our model, however, the small companion G4 directly contacts
the larger member G1, bringing stronger tidal damage on both galaxies.
As seen in the first snap shots in the two figures, both galaxies have been strongly
disturbed by the collisional encounter; many test particles of the galaxies have been
stripped off or transferred to the other galaxy.
In general, gas tends to be affected more by an encounter than the stars.
As a related difference, many more gas particles from G4 than
star particles are seen to be captured by G1
(compare the top row of Fig.~\ref{fig:5}
to that of Fig.~\ref{fig:6}).
The accreted gas at the central region of G1 during the encounter might play a role
to the development in G1 as an active galaxy (as NGC 7319 has been found
to have a type 2 Seyfert nucleus).

The two arms drawn out of G1 continue to grow through the time
shown in the second rows of the two figures ($t$ = 670~Myr),
forming a parallel configuration. As intended in this model, G2 does not distort
the tails while they develop since it stays far from them; meanwhile, G3 continues
to approach the group.
At the instant of the second snapshots, G3 is about to collide with G2 at $z \sim -70$ kpc
in the adopted scaling.
Here G2 is placed slightly to the west (the positive $x$ direction) of G3 at the onset
of the collision.
This is intended 
to make a partial head-on collision between G2 and G3, and to stabilize the orbit of G3
between G1 and G2, keeping G3 from being attracted too much toward G1 afterward,
even though the effect would be small due to the high $z$-speed of G3.
Note that we are not completely certain whether this collision between G2 and G3 does
in fact occur (and in the manner described in the model).
We attempt the collision at the second stage in the model based on
the interpretation (described in Section \ref{sec:2.2}) that some of the disturbed features
observed around the pair NGC 7318a/b could be the result of the earlier interaction
between the two galaxies.
It is thought that
some features, such as SW-arm and stripped gas of G3, might have tidal origin that require
some time (a few 100 Myr) to develop to the presently observed state.

The third and bottom rows of the two figures (Figs~\ref{fig:5} and \ref{fig:6})
show the model system at the times near present,
when the high-speed intruder is passing the mid-plane of G1
colliding with mainly the outer tail and some other particles placed west of G1
($t$ = 900 Myr $\equiv{t_3}$; between shortly before the present and the present)
and when the intruder has passed the mid-plane of G1 after the strong impact
($t$ = 1080 Myr $\equiv{t_4}$; between the present and shortly after the present), 
respectively.
Since G3 collided with G2 earlier it has been evolving tidal structures.
G3 in this model develops moderate tidal tails by the onset of the current collision,
so the disc and the eastern tidal arm 
pulled out of it both hit the IGM at a high-speed
(the $z$-velocities of the model galaxies
will be discussed in Section \ref{sec:4.3}).

The morphology of the model at $t_3-t_4$
looks generally similar to that of SQ.
First of all, the positions of the four model galaxies projected on to the $x$-$y$ plane are
close to those observed.
The parallel tails look similar to the observed features as well.
It should be emphasized that even though the inner and outer tails in the model were
generated simultaneously, the inner tail appears stronger than the outer tail in stellar features.
Some of the features around NGC 7318a/b were generated very roughly in the
model: The two tails drawn out of G3 (by the collision with G2) in stars do not match with the
real optical features very well.
However, the one extracted from the eastern side of the disc looks similar to the
feature SW-arm, although the curvature and orientation are not quite right.
A feature like SQ-A is produced at the north of G3;
the feature began to be formed right after the impact and has grown with time
(it is quite small at $t_3$, but larger at $t_4$; see also Fig.~\ref{fig:7}).
Note that a gas bridge is formed (seen in the bottom right panel of Fig.~\ref{fig:6}) 
after the strong impact of G3 with the IGM. 
(We use the term `gas bridge' for the gas feature running manly in the 
north-south direction formed in the shock region between G1 and G3 
and distinguish it from `H$_2$ bridge' which indicates specifically 
the thin dusty bridge running 
in the east-west direction shown in Fig.~\ref{fig:3}(b).) 
We will discuss and compare the gas structure more in Section \ref{sec:4.3}.

\subsection{Models A versus models C}
\label{sec:4.2}

The fiducial model described above is one of the models 
of type~A (which are designed to generate
the inner and outer tails simultaneously by an encounter of G4 with G1 as suggested
in scenario A in Section \ref{sec:2.2}).
The observations of the inner and outer tails in SQ suggest a younger age of the inner tail
than the outer tail (e.g., \citealt{Fedotov2011}). 
%(e.g. the inner tail is optically brighter, bluer, and thinner). 
We thought at the beginning of this work that models C (which intend to generate the
outer tail first by an earlier encounter of G4 with G1 and then the inner tail by the recent
encounter of G2 with G1 as suggested in scenario C) would be more appropriate for the system.
However, as we proceeded with the simulations, we found difficulties with models of type C.

First of all, reconstructing two long tails one after one in a good configuration following
scenario C was much more complicated than we expected.
In such models, producing the outer tail first from G1 by an encounter of G4 and letting
G4 leave toward east (to its current position) are easy, however, because the scenario
assumes the inner tail is formed sufficiently later than the outer tail,
G2 has to encounter G1 closely (to produce the inner tail)
\emph{after the previously generated outer tail has already been grown}.
Then, even though G2 could manage to pull the inner tail out of (the already disturbed) G1,
creating a relatively good configuration with the outer tail, G2 at the same time strongly
distorts the outer tail.
This `dilemma' (G2 has to pass G1 closely to produce the inner tail but then it also
destroys the outer tail) occurred in the earlier N-body models as well,
so a very inclined orbit of G2 with which G2 would not disturb the
outer tail was used (see \citealt{Renaud2010} for details).
We applied the similar orbits for G2 in models C, but the effect of G2 on the outer tail
was still too strong due to the extended halo potentials used in the SPH code
(Equation \ref{eq:1}).
To decrease the effect of G2 on the outer tail, we tried particularly smaller cutoffs
for the halo potentials of the model galaxies; the effect was still not eliminated.
(We cut off each halo potential at 2$-$3 times the maximum initial disc
extent for models C, and 5 or more times that for models A.)

Another disadvantage of models C is that it is hard to adjust the orbit of G2 after
it encounters G1 to best fit later developments.
We prefer to place G2 relatively far below at the second stage so it collides with G3
(like in the fiducial model) and so G3 has time to develop tidal features (assuming
some features of NGC 7318b have tidal origins). However, when G2 is placed on an
inclined orbit to make the inner tail without destroying the outer tail at the first stage,
it is difficult to put G2 sufficiently far down at the second stage and then let G2 slow
down in its vertical motion (in $z$-axis) at the third stage 
(as it should have a small $z$-speed relative to G1 near the present).
In addition, apart from the possibility of the collision between G2 and G3 well below
the plane of G1, it is difficult to evolve sufficiently many particles to the western side
of G1 to be hit and shocked by G3 at the third stage in these models.

Models of type C do have some advantages. 
One is that the observed gas distribution
near the parallel tails could also generally be well reproduced. 
Another is that a transient feature
like the observed H$_2$-bridge could be relatively well 
generated by the encounter of G2 with G1 in models~C.
We, however, have not tried to optimize models C as thoroughly as with models A.
Thus, although they do not look promising,
we cannot entirely rule out the models of scenario C.

%fig 7

\subsection{Gas properties}
\label{sec:4.3}

We first compare the kinematical and structural features of gas formed in the fiducial
model at $t_4$ (the time of the bottom rows of Figs~\ref{fig:5} and \ref{fig:6})
to the observed \HI features in three different velocity ranges 
(Fig.~\ref{fig:2}; \citealt{Williams2002}).
At that time, the $z$-velocities of the centres of the model galaxies G2, G3, and G4
(relative to G1) are $-$3.6, 64.7, and $-$3.3 in code units, respectively.
(G3 has much higher $z$-velocity than the others, however, 
its relative $z$-speed is not yet comparable to the observed value ($\sim$900 km~s$^{-1}$) 
with the representative scaling.
We found that other models with different $z$-velocities for G3,
ranging from slightly lower to twice higher than that in the fiducial model, have similar results
as long as G3 has a much higher $z$-velocity than the others).
We will hereafter use `$v_z$' for the $z$-velocity of a test particle
in the non-inertial reference frame of G1 (used in all SPH models),
and `$v_r$' for 
%an observed radial velocity 
either an observed radial velocity or 
the velocity of a test particle expressed as the radial velocity. 
To compare with the \HI features in each $v_r$ range,
we roughly divide the gas particles from the model with $v_z$
into three ranges (Fig.~\ref{fig:7}):
$v_z < 20$ (high-$v_r$), $20 \le v_z < 55$ (intermediate-$v_r$), and
$v_z \ge 55$ (low-$v_r$). 
($v_z$= 0 for test particles means the particles have the same $z$-velocities 
to that of the centre of G1; 
positive/negative values of $v_z$ indicate motions toward/away from us relative to G1. 
While, for the observed \HI gas, $v_r$ values less/greater than that of NGC 7319 
($\sim$6600 km~s$^{-1}$) 
mean motions toward/away from us relative to NGC 7319.)      
In contrast to the \HI observation,
where almost no gas was detected in the central regions of all members,
our models (both types of models A and C) did not remove most of the gas
from the members.
(Specifically, the percentage of the gas particles of G1 remaining within
the initial stellar disc radius of G1 at $t_4$ is $\sim$39 per cent 
in number of particles; 
similarly, the percentages of the remaining gas particles of G2, G3, and G4
are about 96 per cent, 45 per cent, and 49 per cent, respectively.)
However, some other structural features of gas were reproduced
relatively successfully in the fiducial model,
as described in the following paragraphs.
It is yet unclear why the galaxies contain no \HI within them.
Some members appear to be of quite early type, and might lost
their gas long before the recent encounters.
The case of NGC 7319 is more mysterious.
Perhaps the bar has funneled much gas to the centre.
Unveiling the origins of the \HI deficiency in the members
is one of the problems
that we need to keep investigating.

As shown in the top row of Fig.~\ref{fig:7},
the gas corresponding to the high-$v_r$ range consists of particles originating from
mostly G1, G2, and G4 (which have a similar high 
radial velocity themselves) as expected (note that G2 was initially set with the smallest
number of gas particles as indicated in Table \ref{table:1}).
A few particles originating from the high-speed intruder G3 (3.5 per cent of G3 gas particles)
which are captured by the others or scattered into the intergalactic space are in this $v_r$ range.
Many high-$v_r$ gas particles are found between G1 and G4, tracing the stellar parallel tails
(shown in the bottom row of Fig.~\ref{fig:5}).
These groups of gas particles in the model look quite similar to the two huge arc-like \HI clouds
observed along the optical inner and outer tails in high-$v_r$ range
(Arc-N and Arc-S in Fig.~\ref{fig:2}, respectively).
Arc-S extends well beyond the southern tip of the optical outer tail, and the gas feature formed
in the model along the outer tail extends further than the corresponding stellar feature.
Arc-N curves toward north (as the UV inner tail; see Section \ref{sec:2})
rather than toward NGC 7320c. 
In the model many gas particles distributed along the inner tail
seem to be curved toward G4; however, 
there is a trail of gas curved far to the north
like the observed cloud Arc-N.
Another notable high-$v_r$ gas feature is seen in the top-left panel of the figure,
at west of G1 and north of G3.
We interpret this gas as the compact high-$v_r$ \HI cloud 
(NW-HV)
detected near SQ-A, although the model feature
should be separated more from both G1 and G3.
(We tried to refine the feature in some similar models by using different values of $v_z$
and/or tilt for G3. However with the limited resolution, the feature did not come out
noticeably differently.)

Gas particles in the intermediate-$v_r$ range
(middle row of Fig.~\ref{fig:7}) appear mostly at north of G3 (left panel).
Most of these gas have come from either G1(35 per cent)
or G3 (63 per cent)
(note that the percentages vary 
depending on the criteria for dividing the $v_r$ ranges;
it is also affected by the number of gas particles set in each member).
The observed intermediate-$v_r$ \HI cloud (NW-LV)
coincides with the high-$v_r$ cloud (NW-HV) near SQ-A,
but more extended.
The intermediate-$v_r$ gas particles in the model are distributed
over a larger area than the high-$v_r$ gas particles at north of G3.

The gas particles in the low-$v_r$ range are presented
in the bottom row of the figure.
Almost all (99 per cent) of the particles originate from G3
(87 per cent of G3 gas particles are in this range).
The \HI observation found low-$v_r$ gas in south of
NGC 7318a/b diffusely distributed
(the feature SW in Fig.~\ref{fig:2}),
and no gas at the central area of NGC 7318b.
As pointed out earlier, in contrast to observation,
many gas particles in the model remain in the disc of G3 and
the spiral features look somewhat different.
However, some of the particles found south of G3 appear
more or less similar to the diffuse \HI gas.

\begin{figure}
\centering
\includegraphics[width=7cm]
{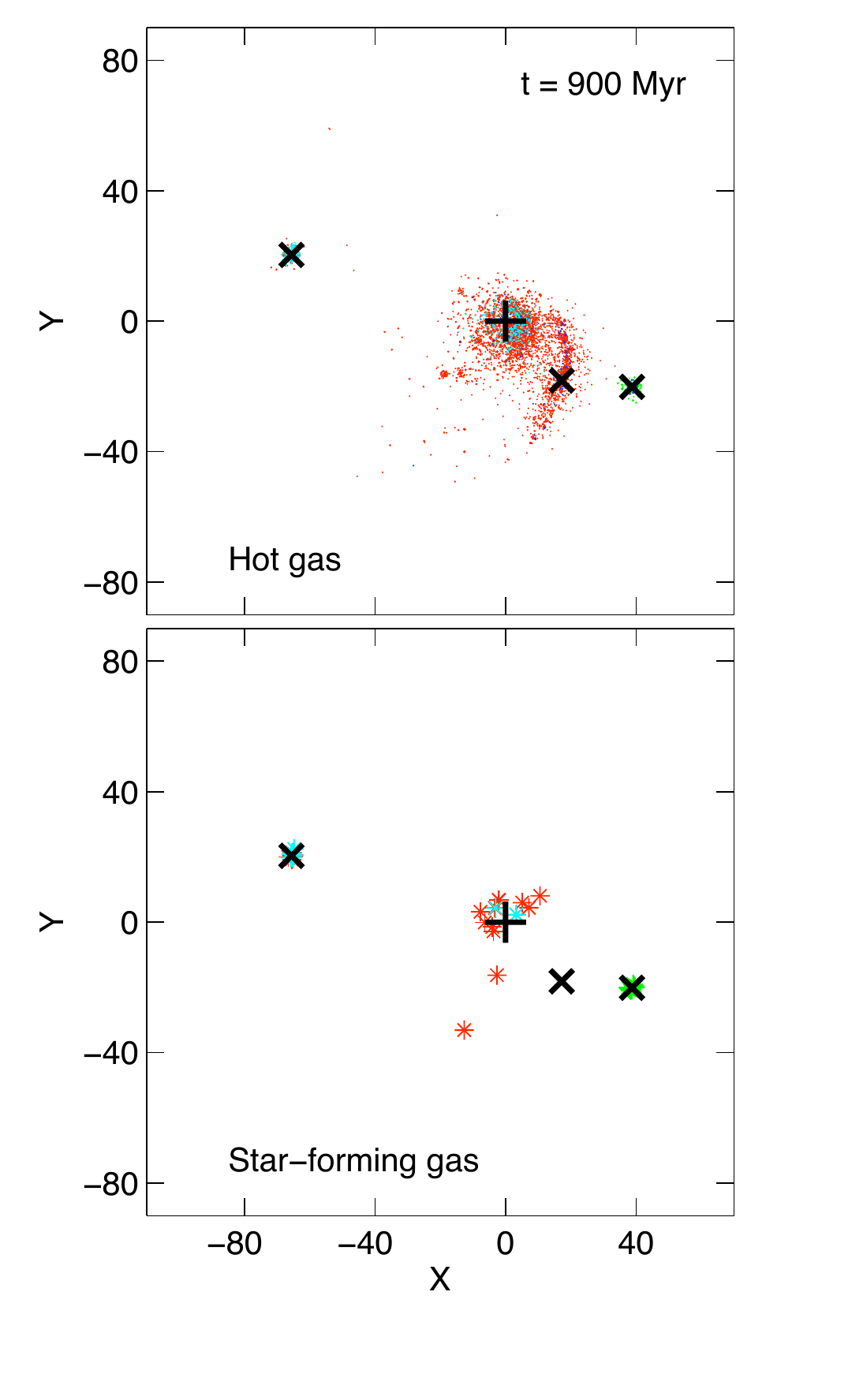} \\ [-20pt]
\caption
{
Hot gas 
and star-forming gas particles 
of the fiducial model at $t$ = 900 Myr. 
In the top panel, gas particles exceeding a temperature cutoff
(see the text) and in the high-$v_r$ range are displayed with red, green,
blue, and cyan dots (for the particles originating from
G1, G2, G3, and G4, respectively).
The centre position of G1 at the instant is marked with 
a plus sign and 
those of G2, G3, and G4 with `X's.
In the bottom panel, star-forming gas particles in the high-$v_r$ range
at the time 
are shown with asterisks.
\label{fig:8}
}
\end{figure}

Next we briefly describe the hot gas distribution 
from the fiducial model when G3 hits the IGM.
The particles shown in the top panel in Fig.~\ref{fig:8}
are those 
hot gas (which is more like H$\alpha$ gas than hot X-ray gas) 
exceeding a certain (arbitrarily chosen)
common temperature cutoff, in the high-$v_r$ range at $t_3$ 
(about 17 per cent of the total gas particles have higher temperatures
than the cutoff at the time, and it decreases to about 6.6 per cent
at about 0.35 time unit (70 Myr) before and after $t_3$
with the same cutoff).
In the bottom panel of the figure, we display
the star forming gas particles
of the model at $t_3$
to compare the distribution of the 
hot gas 
to that of the star forming gas.
Even though the model results in the figure are too rough
to be able to specifically compare to observations,
due to the simplified treatment of heating and cooling in our code,
we can see an elongated feature of hot gas particles (top panel),
with little star formation in the area of the elongated feature (bottom panel).
This implies that the elongated feature in the model
would be heated by the collision with G3 rather than
by star formation, in agreement with the observations.
The model does not provide any insights concerning the structure and 
formation mechanisms of two major star formation 
complexes at either end of the shock structure.

%fig 8

%fig 9
\begin{figure}
\centerline{
\includegraphics[width=7cm]{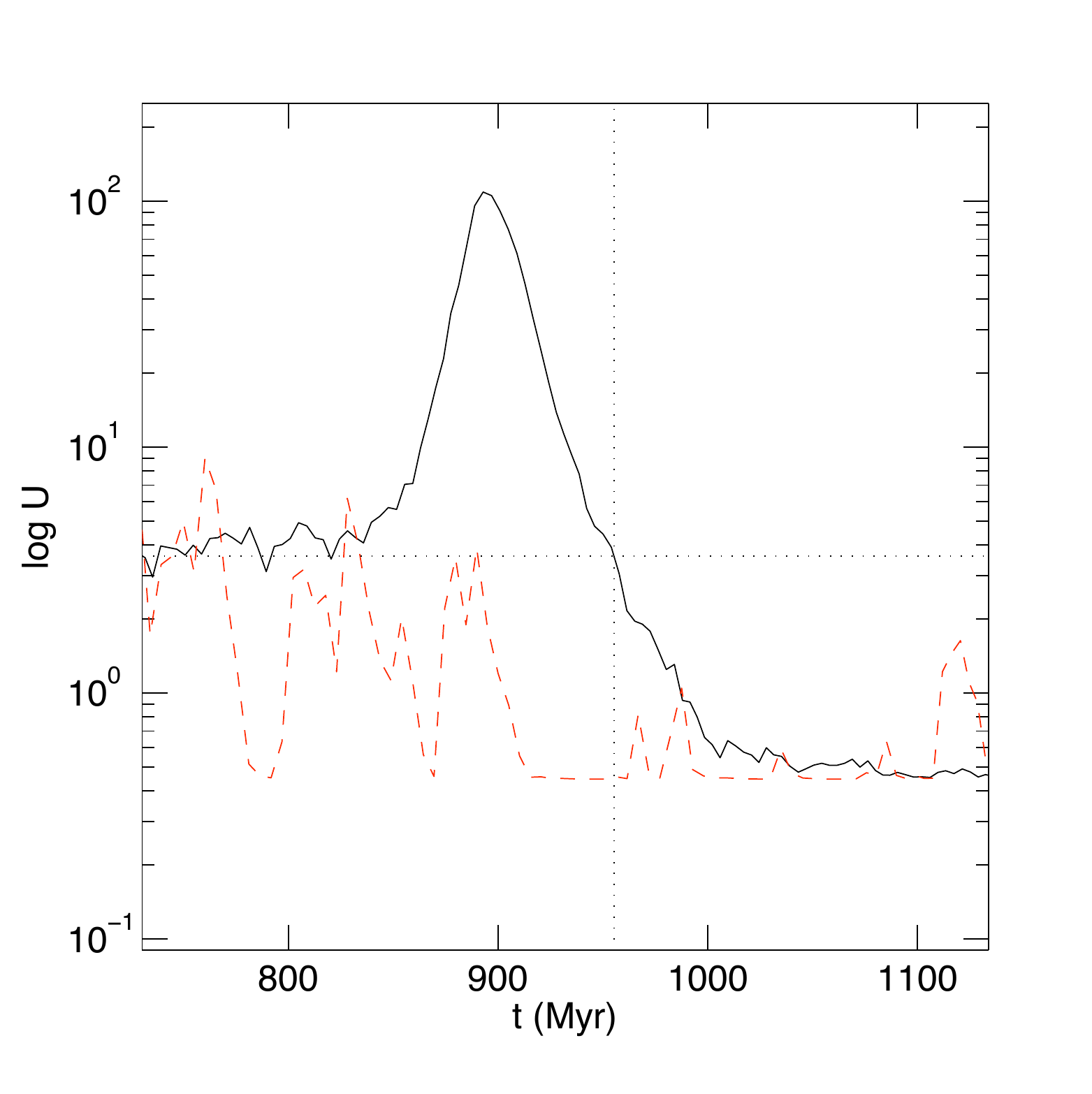}}
\caption
{
The change of the average internal energy of the gas particles in the shock region
in the fiducial model with respect to time (black solid line). 
The change in a comparison model, in which no gas particles of G3 have been set,
is also presented (red dashed).
(See text for the selection criteria of the gas particles considered 
for the internal energies per particle in both models.)
The time on the $x$-axis is converted with the representative scaling
(1 time unit = 200 Myr).
G3, the high-speed intruder, passes the mid-plane of the disc of G1 in both models
at about t = 890 Myr in the adopted scaling.
The average internal energy of the gas particles in the fiducial model 
decreases rather slowly after the strong impact, 
having a nearly power-law form, for some tens of millions of years.
Not the absolute values but the relative values of the $y$-axis are meaningful. 
The horizontal dotted line is drawn at the $y$ value of 
the fiducial model at t = 730 Myr, as a representative 
value of the average internal energy before the strong impact; 
the vertical dotted line is drawn at t = 955 Myr when the average value 
starts to go below
the representative value again after the strong impact.
\label{fig:9}
}
\end{figure}

In Fig.~\ref{fig:9},
we show the average of the internal energy  
of the shock-heated gas particles in the fiducial model with respect to time (black solid line).
For the internal energy per particle, 
we include all of the gas particles of G1 and G3
which appear near the shock and the gas bridge formed shortly after the strong
impact of G3 with the IGM in the fiducial model. 
Specifically, the particles within the region of $0 \leq x \leq 30$, $-40 \leq y \leq 10$,
and $10 \leq z \leq 30$ at $t =  5.0$ (1 Gyr in the representative scaling)
in code units were chosen. 
The centre position of G3 at the time is about
$x$ = 17, $y$ = $-$20, and $z$ = 40.
To compare to the fiducial run, we built a similar model,
setting no gas particles for G3 and keeping all the other parameters of the fiducial model.
The average internal energy of the gas particles of G1 that appear within the same region
at the same time in the comparison model is also presented in the figure (red dashed line).
(In the fiducial model, total 2779 gas particles, 1447 from G1 $+$ 1332 from G3,
met the above selection-criteria. While, in the comparison model,
170 gas particles of G1 met the same criteria. 
For the internal energies per particle represented in the $y$-axis in the figure, 
these 2779 and 170 gas particles are considered 
in the fiducial and comparison models, respectively.) 
In both models, G3 collides with the IGM at about $t$ = 4.45,
which corresponds to 890 Myr in the adopted scaling.
(G3 in these models collides first with G2 at about $t$ = 3.38, 700 Myr.)
As expected, a strong peak occurs in the fiducial model but not in the comparison one
at about the impact, when G3 hits the outer tail 
and some other stripped materials west of G1.
In the fiducial model, the average internal energy of the gas particles
in the shock region decreases rather slowly after the impact,
showing a nearly power-law form as might be expected from decaying turbulence
in the interacting flows.
We checked the trajectories of some of the gas particles (clouds)
and found that many of the particles' trajectories 
in the gas bridge in the fiducial model
were redirected by collisions with other particles,
after G3 has already passed through the plane of G1,
for some tens of millions of years.
This amounts to decaying turbulence with many small shocks 
of various strengths in that region and we think 
it can account for the observed emission without much star formation.
The time-scale is much longer than the cooling time of
a single impact shock (e.g., \citealt{Guillard2009}),
and so makes it easier to understand how we happen to see 
a phenomena that would otherwise be extremely short-lived.

\section{Summary and discussion}
\label{sec:5}

Motivated by the disturbed structure of SQ, including the remarkable group-wide shock seen
at the IGM between NGC 7319 and 7318b, and using published high quality observations for
the system in various wavebands,
we have attempted to model the interaction history and the evolution of the large-scale structure
of the group.

Despite the complex interactions expected in the compact group,
some of its major features allow us to constrain the models reasonably well.
First, the appearance of the parallel tails provides important clues to track how they
were formed, narrowing down the possibilities to encounters of 
NGC 7320c and/or NGC 7318a with NGC 7319.
In addition, the fact that such delicate tidal features extend large distances outward
justifies the supposition that members might not pass close to the tails after they were
produced, otherwise they would have been destroyed or greatly modified.
Some of the other complex disturbed features around the pair NGC 7318a/b,
such as the tail-like feature SW-arm and the stripped gas disc of the high-speed intruder,
suggest NGC 7318b might have interacted with NGC 7318a before.
At the current time, it is known from many observations that the high-speed intruder,
NGC 7318b, has hit the IGM giving rise to the large-scale shock.

The interactions described above seem to take place one after another,
involving mainly two members at a time; which simplifies the modeling.
Thus we divided each of our models in three major stages and simulated a series of
plausible interactions in order: At the first stage, we tried a collision between NGC 7319
and 7320c to produce both of the parallel tails simultaneously in models (models A),
or via an encounter between NGC 7319 and 7320c to generate the outer tail first and
then an encounter between NGC 7319 and 7318a to pull out the inner tail later (models C).
Then we attempted at the second stage an interaction between NGC 7318a and 7318b of
varying intensity, and at the final stage a collision between NGC 7318b 
with the IGM.

Models A were generally more successful than models C. Even though generating the parallel
tails one by one from two different interactions seems to be conceptually reasonable,
in models C it turned out to be very difficult to pull another substantial tail (the inner tail)
in a second close encounter, while at the same time preserving the outer tail.
This later encounter usually destroyed the outer tail in models with extended halo potentials,
and with very limited haloes it still disturbed the outer tail, yielding a poor tail morphology.
On the other hand, a single strong encounter, as in models A, could generate two substantial
tails nicely with different stellar intensity. 
\emph {We note that the fact that the two tails don't have 
the same star formation history 
doesn't have to mean that they weren't born in the same tidal event.}

In the fiducial model (one of the best models of type A),
the inner and outer tails are generated simultaneously
by an encounter of NGC 7320c,
and the high-speed intruder experiences an earlier collision with NGC 7318a,
the galaxy slightly to the west of it, below the plane of NGC 7319.
This early collision induces spiral waves in the intruder;
one of which later collides with the outer tail and the IGM.

We think that
the fiducial model is generally successful
at reproducing the large-scale morphology and kinematics of SQ.
Specifically, the current relative positions (projected on to the sky plane)
and relative radial velocities of the members,
the long parallel tails with thinner and brighter inner tail in stellar feature,
the huge amount of high-$v_r$ gas along both of the parallel tails,
the features looking like SW-arm and SQ-A, the high and intermediate-$v_r$ gas
at north of the high-speed intruder and some scattered low-$v_r$ gas at
southwest of the high-speed intruder that were produced in the model
resemble more or less the real features. 
However, the detailed inner structures of the spiral members, 
the many disturbed structural features around NGC 7318a/b, 
and the gas removal from each disc (which may also resulted from the earlier interactions 
than our simulations start) were not reproduced in our models. 
The location of SQ-A and the orientation of SW-arm 
in the model do not match the observations well either. 
The gas temperature and star-formation history of our models
are not accurate enough for detailed comparison to observations.
The shortcomings of the models are the result of limited resolution and the approximate
treatments of heating and cooling in the current code.
Nonetheless, the model shows the shock-heated gas, by the impact of the high-speed
intruder with the IGM, in an elongated feature with little star formation.
A gas bridge is formed in the shock region 
and particles in the bridge continue to interact for some
tens of millions of years after the impact.
Dynamical friction has been neglected 
in this work. Though the effect would not change our main results, 
it becomes particularly significant for modeling the future of the system.

Finally, we note some general implications from this work.
Constructing numerical models of a compact group could be very difficult
as the system might have experienced multiple interactions.
However, if a group possesses very extended features,
the model could be relatively well constrained.
This is because not only do the rich features themselves provide
important clues to the formation, developing and maintaining
those delicate extended features also could eliminate the possibilities of
strong interactions occurring in the vicinity of the features after the production.
We found a very interesting result in SQ, with collisions occurring 
mostly two-at-a-time.
This result may be general (although it is not necessarily so),
because flyby or the final merger together time would be short compared to
the total orbital time.

SQ appears to be unusual in the population of compact groups 
in that the system possesses prominent, extended features 
and the intruder seems to be actively colliding with 
previously known tidal material. 
This might be due to the unique interaction history of the system, 
or result from the short time-scale during which such features can be observed. 
SQ seems to exhibit a wide variety of galaxy interaction indicators, 
ranging from clear stellar and gaseous tidal features to group-wide shock structures. 
This kind of diversity may be special to the compact group environment, 
but SQ may be a snapshot in time of what many compact groups may go through 
if we wait long enough. 
The question of how commonly powerful shocks are found 
in the compact group environment is being explored by a major \emph{Spitzer} survey
of HCGs by Cluver el al. (in preparation). 
We expect that our major result can be extended to other similar groups in
studying the evolutionary histories or finding initial conditions of collisions.
Detailed numerical studies on more compact groups would also
help to better understand the evolution of high-redshift galaxies.

\section*{Acknowledgments}
J.-S.H. thanks Korea Institute for Advanced Study for providing computing resources
(KIAS Center for Advanced Computation Abacus System) 
for the part of this work.

\bsp

\label{lastpage}

\end{document}